\documentclass[10pt,prl,aps,twocolumn,superscriptaddress,showpacs,nofootinbib]{revtex4-1}
\usepackage{amscd}
\usepackage{amsthm,amsfonts,dsfont}
\usepackage{enumerate}
\usepackage{float}

\usepackage{svg}
\usepackage{mathrsfs}
\usepackage{amssymb}
\usepackage{graphicx}
\usepackage{dcolumn}
\usepackage{bm}
\usepackage{textcomp}
\usepackage{amsmath}
\usepackage[titletoc]{appendix}
\usepackage[hidelinks]{hyperref}
\usepackage{epsfig}
\usepackage{epstopdf}
\usepackage{subfigure}
\usepackage{color}
\usepackage[10pt]{moresize}
\usepackage{soul}

\newcommand\ket[1]{\ensuremath{|#1\rangle}}

\newcounter{RomanNumber}

\begin{document}

\title{Bayesian Phase Stabilization at the Shot‑Noise Limit for Scalable Quantum Networks}

\affiliation{Hefei National Research Center for Physical Sciences at the Microscale and School of Physical Sciences, University of Science and Technology of China, Hefei, Anhui 230026, China}
\affiliation{Hefei National Laboratory, University of Science and Technology of China, Hefei, China}
\affiliation{Jinan Institute of Quantum Technology and CAS Center for Excellence in Quantum Information and Quantum Physics, University of Science and Technology of China, Jinan 250101, China}
\affiliation{Shanghai Research Center for Quantum Science and CAS Center for Excellence in Quantum Information and Quantum Physics, University of Science and Technology of China, Shanghai, China}

\author{Guang-Cheng Liu}
\affiliation{Hefei National Research Center for Physical Sciences at the Microscale and School of Physical Sciences, University of Science and Technology of China, Hefei, Anhui 230026, China}
\affiliation{Hefei National Laboratory, University of Science and Technology of China, Hefei, China}
\affiliation{Shanghai Research Center for Quantum Science and CAS Center for Excellence in Quantum Information and Quantum Physics, University of Science and Technology of China, Shanghai, China}

\author{Chao-Hui Xue}
\affiliation{Hefei National Research Center for Physical Sciences at the Microscale and School of Physical Sciences, University of Science and Technology of China, Hefei, Anhui 230026, China}
\affiliation{Hefei National Laboratory, University of Science and Technology of China, Hefei, China}
\affiliation{Jinan Institute of Quantum Technology and CAS Center for Excellence in Quantum Information and Quantum Physics, University of Science and Technology of China, Jinan 250101, China}

\author{Fa-Xi Chen} %
\author{Ming-Yang Zheng}%
\affiliation{Hefei National Laboratory, University of Science and Technology of China, Hefei, China}
\affiliation{Jinan Institute of Quantum Technology and CAS Center for Excellence in Quantum Information and Quantum Physics, University of Science and Technology of China, Jinan 250101, China}

\author{Yi Yang}
\affiliation{Hefei National Research Center for Physical Sciences at the Microscale and School of Physical Sciences, University of Science and Technology of China, Hefei, Anhui 230026, China}
\affiliation{Hefei National Laboratory, University of Science and Technology of China, Hefei, China}
\affiliation{Jinan Institute of Quantum Technology and CAS Center for Excellence in Quantum Information and Quantum Physics, University of Science and Technology of China, Jinan 250101, China}

\author{Li-Bo Li}%
\affiliation{Jinan Institute of Quantum Technology and CAS Center for Excellence in Quantum Information and Quantum Physics, University of Science and Technology of China, Jinan 250101, China}

\author{Bin Wang} %
\affiliation{Jinan Institute of Quantum Technology and CAS Center for Excellence in Quantum Information and Quantum Physics, University of Science and Technology of China, Jinan 250101, China}

\author{Bo-Wen Yang}
\author{Hai-Feng Jiang}%
\author{Yong Wan}
\author{Ye Wang}
\affiliation{Hefei National Research Center for Physical Sciences at the Microscale and School of Physical Sciences, University of Science and Technology of China, Hefei, Anhui 230026, China}
\affiliation{Hefei National Laboratory, University of Science and Technology of China, Hefei, China}
\affiliation{Shanghai Research Center for Quantum Science and CAS Center for Excellence in Quantum Information and Quantum Physics, University of Science and Technology of China, Shanghai, China}

\author{Jiu-Peng Chen}
\affiliation{Hefei National Laboratory, University of Science and Technology of China, Hefei, China}
\affiliation{Jinan Institute of Quantum Technology and CAS Center for Excellence in Quantum Information and Quantum Physics, University of Science and Technology of China, Jinan 250101, China}

\author{Qiang Zhang}
\affiliation{Hefei National Research Center for Physical Sciences at the Microscale and School of Physical Sciences, University of Science and Technology of China, Hefei, Anhui 230026, China}
\affiliation{Hefei National Laboratory, University of Science and Technology of China, Hefei, China}
\affiliation{Jinan Institute of Quantum Technology and CAS Center for Excellence in Quantum Information and Quantum Physics, University of Science and Technology of China, Jinan 250101, China}
\affiliation{Shanghai Research Center for Quantum Science and CAS Center for Excellence in Quantum Information and Quantum Physics, University of Science and Technology of China, Shanghai, China}

\author{Jian-Wei Pan}
\affiliation{Hefei National Research Center for Physical Sciences at the Microscale and School of Physical Sciences, University of Science and Technology of China, Hefei, Anhui 230026, China}
\affiliation{Hefei National Laboratory, University of Science and Technology of China, Hefei, China}
\affiliation{Shanghai Research Center for Quantum Science and CAS Center for Excellence in Quantum Information and Quantum Physics, University of Science and Technology of China, Shanghai, China}

\begin{abstract}
High-precision optical phase stabilization in quantum networks is fundamentally constrained by the strict photon-flux and duty-cycle limits required to avoid disturbing fragile quantum states. This challenge becomes especially critical when coordinating multiple independent light sources for multi-step quantum protocols. Here, we develop an integrated phase-stabilization framework that incorporates a Bayesian phase estimator to optimally extract information from sparse single-photon detection events. This approach outperforms conventional maximum-likelihood estimation and achieves the shot-noise limit under minimal photon flux. The framework enables real-time correction of combined phase noise from both nodal lasers and transmission fibers, facilitating a two-step excitation protocol for heralded entanglement generation between separate trapped-ion nodes via single-photon interference. Operating with a detected photon rate of $\sim$1~MHz and a duty cycle $\leq 6.5\%$, the system maintains interferometric visibility $>97\%$ over fiber links of 10 km and 100 km. This phase control yields long-lived ion-ion entanglement with parity contrast exceeding $85\%$, enabling device-independent quantum key distribution with finite-size security analysis over 10 km and a positive asymptotic key-rate over 100 km. Moreover, the resulting memory-memory entanglement at 10 km survives beyond the average time required to establish it—a fundamental requirement for quantum repeaters. This work establishes a robust and scalable foundation for practical long-distance quantum networks.
\end{abstract}

\maketitle

{\it Introduction.---}
High-precision optical phase coherence between distant nodes is essential for large-scale quantum networks~\cite{kimble2008quantum,duan2010colloquium,azuma2023quantum}, as it enables high-visibility single-photon interference for efficient interconnection~\cite{cabrillo1999creation,duan2001long,lucamarini2018overcoming}. A prevailing strategy to establish and maintain such coherence involves interleaving faint reference pulses with quantum signals for active phase tracking~\cite{minder2019experimental,wang2019beating,liu2019experimental,fang2019surpassing,yu2020entanglement,van2022entangling,stolk2024metropolitan,jabir2025enabling,liu2026long,lu2026device}. This technique has substantially advanced all-optical quantum secure communication systems, notably achieving twin-field quantum key distribution~\cite{lucamarini2018overcoming} over distances exceeding 1000 km~\cite{liu2023experimental}. However, extending these methods to matter‑based nodes—such as trapped ions~\cite{PhysRevLett.124.110501,nadlinger2022experimental,main2025distributed}, single atoms~\cite{van2022entangling,zhang2022device}, atomic ensembles~\cite{yu2020entanglement,liu2024creation}, or solid-state defects~\cite{stolk2024metropolitan,knaut2024entanglement,hermans2022qubit}—presents unique challenges. 

Protocols employing matter nodes typically involve intricate pulse sequences for qubit manipulation, which severely restrict the optical duty cycle and force any interleaved reference pulses to be temporally sparse to avoid qubit decoherence~\cite{yu2020entanglement,van2022entangling,stolk2024metropolitan,liu2026long,lu2026device}. Operating with these references at the single-photon level—a common practice in all-optical systems—imposes a stringent photon‑budget constraint on phase estimation. In such photon-starved regime, conventional maximum‑likelihood estimation (MLE)~\cite{hacker2023phase} faces a fundamental precision–delay trade‑off: rapid sampling is limited by shot noise, while extended integration accumulates unbounded phase diffusion from environmental disturbances. This challenge is acute in multi-step protocols~\cite{maurer2004single,liu2026long} that must track combined phase evolution from multiple independent lasers while preserving qubit coherence. As a result, current implementations involving matter nodes often resort to using bright classical references~\cite{yu2020entanglement,van2022entangling,stolk2024metropolitan,stolk2025extendable}. Although this alleviates photon‑statistical limits, the requisite strong optical isolation introduces additional phase noise through auxiliary paths, ultimately limiting tracking accuracy and hindering overall scalability. 

In this work, we demonstrate an integrated phase stabilization framework that incorporates a Bayesian phase estimator. Our estimator optimally combines prior knowledge of phase-diffusion dynamics with sparse photon‑detection events, enhancing the effective Fisher information (FI) beyond that of conventional MLE. The framework actively corrects combined phase noise from both nodal lasers and transmission fibers, enabling a two-step excitation protocol for heralded entanglement generation between separate trapped-ion nodes via single-photon interference. Operating with weak phase‑reference probes ($\sim$1 MHz photon flux) at low optical duty cycle ($\leq6.5\%$), the system maintains shot-noise limited performance with $>97\%$ interferometric visibility over 10 km and 100 km fiber links. This phase control yields deterministic ion-ion entanglement with parity contrast exceeding $85\%$ at both distances, enabling device-independent quantum key distribution~\cite{zhang2022device,nadlinger2022experimental,liu2022toward,liu2026long,lu2026device}. Moreover, the resulting memory-memory entanglement at 10 km persists longer than the average time required to establish it—a fundamental requirement for quantum repeaters~\cite{briegel1998quantum}.

{\it Bayesian Phase Estimation with FI Advantage.---}
We analyze phase tracking for a time-varying phase $\phi(\tau)$ in a Mach-Zehnder interferometer probed by weak coherent states. The mean photon counts at the output ports over integration time $\tau$ follow $\lambda_{1,2}(\phi) = \frac{N}{2}\left[1 \pm V_0\cos\phi\right],$ where $N = \mu\tau$ is the mean photon number, $\mu$ is the photon flux, and $V_0$ is the maximum interference visibility. Poisson-distributed counts $\lambda_{1,2}$ yield classical FI:
\begin{equation}
I_{\mathrm{F}}(\phi) = \frac{N V_0^2 \sin^2\phi}{1 - V_0^2 \cos^2\phi}.
\end{equation}
At $\phi = \pi/2$, $I_{\mathrm{F}} = \mu V_0^2 \tau$, setting the shot-noise limit (SNL) for static phase measurement.

Phase diffusion, modeled as a Wiener process with $\langle[\delta\phi(\tau)]^2\rangle = D \tau$, fundamentally alters this scaling. Temporal averaging over $\tau$ reduces effective visibility to $V(\tau) = V_0 e^{-D\tau/2}$. The average FI per measurement is then $\bar{I}_{\mathrm{F}}(\tau) = \mu V_0^2 \tau e^{-D\tau}.$

Consequently, the estimation variance for conventional MLE is bounded by
\begin{equation}
    \sigma_{\rm Conv}^2(\tau)\gtrsim e^{D\tau}/(\mu V_0^2\tau)+D\tau,
\end{equation}
where the first term is the measurement error and the second is the diffusion-induced error. Minimizing over $\tau$ reveals an inescapable precision-delay trade-off.

We overcome this limitation by incorporating prior knowledge within a recursive Bayesian framework. Under Gaussian assumptions for both prior and likelihood, the estimation variance evolves as~\cite{valeri2020experimental}
\begin{equation}
\sigma_{k+1}^{-2} = \left(\sigma_k^2 + D\tau\right)^{-1} + \Gamma\tau e^{-D\tau},
\end{equation}
where the prior variance \(\sigma_k^2\) first degrades by diffusion and is then updated with the measurement FI. To ensure robustness against outliers, we implement a nonlinear innovation filter with a threshold multiplier \(\kappa\), characterized by an efficiency \(\eta(\kappa)\approx 1\) for typical \(\kappa\in[1,1.5]\). The steady-state variance then reaches the tracking SNL
\begin{equation}
\sigma_{\infty}^2 \approx \sqrt{\frac{D}{\eta(\kappa)\Gamma}} = \left(\frac{D}{\eta(\kappa)\mu V_0^2}\right)^{1/2}.
\label{eq:SNL_tracking}
\end{equation}

The prior-assisted FI, combining the stabilized prior and new data, is 
\(I^{\mathrm{prior}}_{\mathrm{F}} = 1/(D\tau+\sigma_{\infty}^2)\). The corresponding estimation variance with prior assistance is bounded by
\begin{equation}
\sigma^2_{\mathrm{Bay}}(\tau) \gtrsim \frac{1}{I^\mathrm{prior}_{\mathrm{F}}+\bar{I}_F}+D\tau.
\label{eq:error_prior_fast}
\end{equation}
Equation~\eqref{eq:error_prior_fast} reveals that persistent prior information breaks the conventional precision–delay trade-off. For the optimal choice of the shortest possible \(\tau\), the steady-state variance scales as \(\sigma^2_{\mathrm{Bay}}\propto \mu^{-1/2}\)—the hallmark of the SNL—in stark contrast to the \(\sigma^2_{\mathrm{Conv}}\propto \mu^{-1}\) scaling of conventional MLE. While MLE variance diverges as \(\tau\to0\), the Bayesian estimator remains stable and achieves optimal SNL performance.

\begin{figure}[ht!]
 \centering
 \includegraphics[width=\linewidth,trim=0 81 0 110,
 clip]{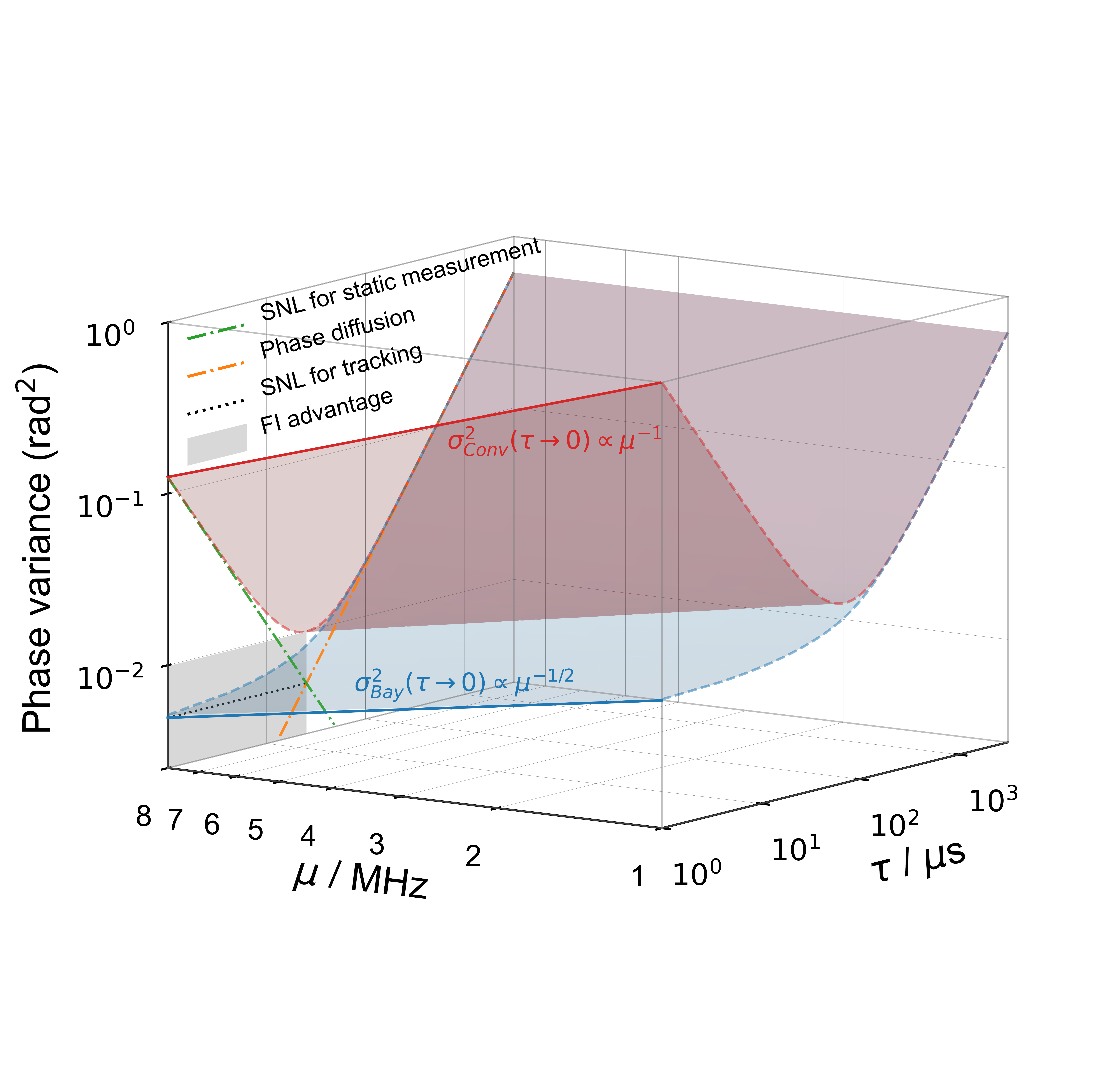}
 \caption{\textbf{Phase stabilization performance comparison.}
 Three‑dimensional representation of the residual phase variance as a function of measurement duration $\tau$ and photon flux $\mu$ for conventional MLE (pink surface) and prior‑assisted Bayesian estimation (blue surface). The orange dotted curve represents the phase‑diffusion noise floor ($\propto D\tau$), and the pink dotted curve indicates the photon shot‑noise limit ($\propto (\Gamma\tau)^{-1}$), which sets the SNL for a static measurement. The black dashed curve shows the SNL for tracking a diffusing phase. The red solid line traces the minimum variance of conventional MLE, which scales as $\mu^{-1}$ and diverges as $\tau \to 0$. In contrast, the blue solid line traces the Bayesian estimator’s variance, which saturates at the SNL scaling $\mu^{-1/2}$ as $\tau \to 0$. The gray shaded region defines the FI-advantage regime, where Bayesian estimation outperforms conventional MLE, effectively mitigating the precision–delay trade‑off and enabling robust phase stabilization across operational bandwidths.}
 \label{fig:Fig2_simu}
\end{figure}

As illustrated in Fig.~\ref{fig:Fig2_simu}, the Bayesian estimator achieves a lower phase variance than conventional MLE. This advantage is most pronounced in photon‑starved regimes where prior knowledge optimally supplements sparse measurement data. The recursive Bayesian estimator, stabilized by outlier rejection, maintains near‑optimal performance across operational bandwidths, enabling robust high‑fidelity phase stabilization (Supplemental Material~\cite{suppmat} S1). Although our analysis assumes a Wiener diffusion model, the Bayesian framework extends naturally to power-law scaling noise while retaining its qualitative advantage over conventional MLE (Supplemental Material~\cite{suppmat} S2.2).

\begin{figure*}[ht!]
\centering
\includegraphics[width=18cm]{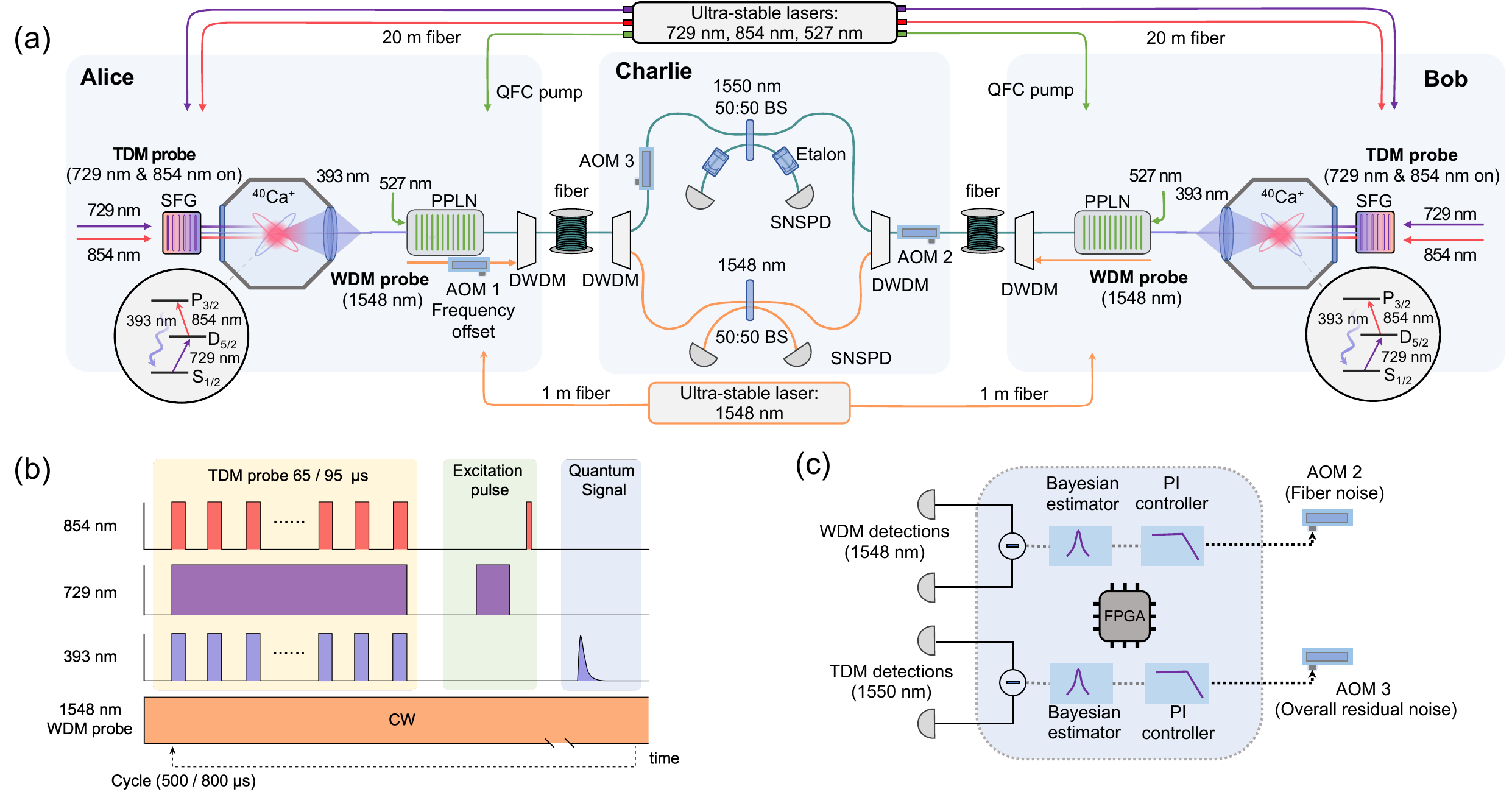}
\caption{\textbf{Experimental setup.} 
\textbf{(a)} Optical layout. Two trapped-ion nodes (Alice and Bob) emit 393-nm photons that are collected by high-NA objectives, frequency-converted to 1550 nm via PPLN waveguides, and transmitted through fiber spools to a central beam splitter at Charlie, where single-photon interference heralds entanglement. Qubit-control light (729 nm and 854 nm) and pump light (527 nm) are delivered to each node through separate 20-m fibers. A continuous 1548-nm reference, multiplexed with the quantum channel, stabilizes fiber-link phase fluctuations (WDM). Weak 393-nm pulses, generated via sum-frequency generation (SFG) from the same 729-nm and 854-nm lasers and interleaved during dead times, compensate residual phase noise (TDM). An etalon filters QFC noise; AOM 1 ensures identical optical frequencies for the 1548-nm references from both arms. AOM, acousto-optic modulator; DWDM, dense wavelength-division multiplexing; SNSPD, superconducting nanowire single-photon detector; QFC, quantum frequency conversion; PPLN, periodically poled lithium niobate. 
\textbf{(b)} Timing diagram. For 10-km (100-km) fiber links, the cycle duration is 500 $\mu$s (800 $\mu$s), with a 65-$\mu$s (95-$\mu$s) window for 393-nm TDM phase-probe pulses (100-ns period, 50\% duty cycle). The 1548-nm WDM reference operates continuously. 
\textbf{(c)} Feedback architecture. Photon counts from WDM and TDM channels are processed by an FPGA executing Bayesian estimation and proportional-integral (PI) control. AOM 2 corrects fiber-induced phase fluctuations (WDM); AOM 3 corrects residual phase noise (TDM).}
\label{fig:setup} 
\end{figure*}

{\it Experiment.---} The integrated phase-stabilization framework, as shown in Fig.~\ref{fig:setup}a, features a dual-band architecture~\cite{pittaluga2021600} incorporating two parallel Bayesian phase estimators to actively correct combined phase noise arising from both the nodal lasers and the long-haul transmission fibers. This architecture enables a two‑step excitation protocol for heralded entanglement generation via single‑photon interference~\cite{cabrillo1999creation,duan2001long} between two independent trapped‑ion nodes (Alice and Bob). Each node traps a single \(^{40}\)Ca\(^+\) ion in a blade trap, with a 3-m free-space baseline between nodes. The 729-nm and 854-nm qubit-control beams, together with the 527-nm frequency-conversion pump, are derived from common, ultra-stable laser sources with Hz-level linewidths. To emulate independent laser systems at the two nodes, these beams are delivered through separate 20-m fibers exposed to ambient acoustic and thermal fluctuations, which impart uncorrelated phase noise between the nodes (Supplemental Material~\cite{suppmat} S2.6).

The two‑step excitation protocol is executed synchronously at both nodes. Each cycle begins with extended Doppler and electromagnetically‑induced-transparency cooling (>$100\ \mu\text{s}$) to reduce atomic‑recoil errors, followed by optical pumping to initialize the ion in the $\ket{S_{1/2}, m_J = +1/2}$ state. A resonant $\pi$‑pulse at 729 nm then prepares the metastable $\ket{D_{5/2}, m_J = +5/2}$ state. A subsequent nanosecond pulse at 854 nm excites the ion to $\ket{P_{3/2}, m_J = +3/2}$ with probability $\alpha$. Spontaneous decay from the $P_{3/2}$ state emits a 393 nm photon, generating an ion‑photon entangled state of the form~\cite{PhysRevLett.110.083603}
\begin{equation}
\ket{\Psi} = \sqrt{1-\alpha}\ket{D,0} + \sqrt{\alpha}e^{i\phi_{\text{oper}}}\ket{S,1},
\end{equation}
where $\phi_{\text{oper}}$ is the relative phase between the 729 nm and 854 nm laser fields imprinted on the photon during the coherent excitation‑and‑decay process.

The emitted 393‑nm photons are collected by high‑numerical‑aperture objectives(NA$=0.635$), coupled into single‑mode fibers (SM300), and directed to a quantum‑frequency‑conversion (QFC) stage. There, the 393 nm photons are converted to the telecom C‑band (1550 nm) via difference‑frequency generation in a periodically‑poled lithium niobate (PPLN) waveguide, pumped by a 527 nm laser~\cite{yang2026high}. The telecom photons from both nodes are transmitted through variable‑length fiber spools (5 km or 50 km) to a central 50:50 fiber beam splitter at station Charlie. A successful Bell‑state measurement, identified by one and only one detection event across two complementary output ports using superconducting nanowire single‑photon detectors (SNSPDs), heralds entanglement between the remote ions in the state $\ket{\Psi^{\pm}} = (\ket{DS} \pm e^{i\theta}\ket{SD})/\sqrt{2}$.

The phase \(\theta\) of the resulting entangled state is
\begin{equation}
\theta = \Delta\phi_{\mathrm{oper}} + \Delta\phi_{\mathrm{paths}}.
\end{equation}
where $\Delta\phi_{\mathrm{oper}} = \phi_{\mathrm{oper}}^A - \phi_{\mathrm{oper}}^B$ is the differential operational phase imprinted during the excitation pulses at Alice and Bob, and $\Delta\phi_{\mathrm{paths}} = \Delta\phi_{\mathrm{QFC}} + \Delta\phi_{\mathrm{fiber}}$ is the total path phase, comprising the phase difference of QFC pump ($\Delta\phi_{\mathrm{QFC}}$) and the differential fiber propagation phase ($\Delta\phi_{\mathrm{fiber}}$). Simultaneous stabilization of both contributions is therefore essential for maintaining entanglement fidelity.

The extended duration of the experimental sequence (Fig.~\ref{fig:setup}b) creates a stringent phase‑stabilization challenge, manifesting as a precision–delay dilemma. To preserve the ions' internal states, strong phase‑reference light cannot be injected during active qubit operations, restricting phase probes to experimental dead times (e.g., during extended cooling). This constraint leads to competing noise sources: rapid sampling with a limited number of probe photons is shot‑noise limited, while extended integration times accumulate unbounded phase diffusion. This tension is further exacerbated by the need to track the combined phase evolution of the two independent laser fields (729 nm and 854 nm) used in the two‑step excitation process. To overcome this, we implement a dual-band stabilization architecture that targets distinct noise sources hierarchically using Bayesian estimators, which optimally extract phase information from scarce probe photons (Fig.~\ref{fig:setup}a).

For rapid fiber fluctuations, we inject a continuous‑wave 1548 nm auxiliary reference, attenuated to the single-photon level and multiplexed with the quantum channel via wavelength-division multiplexing (WDM). At Charlie, the reference is demultiplexed, polarization-purified, and interfered at a dedicated 50:50 beam splitter, with detection by SNSPDs. To avoid injecting broadband noise into the quantum channel via nonlinear effects such as spontaneous Raman scattering~\cite{wang2017long}, we maintain the reference photon flux at approximately 2 MHz per detector.

For the overall residual phase ($\Delta\phi_{\mathrm{oper}}+\Delta\phi_{\mathrm{paths}}$), we implement time‑division multiplexing (TDM) synchronized with the entanglement‑generation sequence. Weak 393 nm phase‑probe pulses (duty cycle $\leq 6.5\%$) are interleaved during experimental dead times, particularly the extended cooling periods. Critically, these probes are generated via sum‑frequency generation in a PPLN waveguide using the identical 729 nm and 854 nm lasers employed for ion manipulation. They subsequently undergo identical QFC with the same 527 nm pump laser, ensuring the probes inherit the exact phase characteristics and noise properties of the quantum signals. At Charlie, the two optical fields are polarization-purified and interfered at a dedicated beam splitter, with detection by SNSPDs. Etalons placed in each output arm suppress nonlinear noise from the QFC process. To prevent re‑Rayleigh scattering from contaminating the quantum signals~\cite{PhysRevLett.124.070501}, we limit the probe photon flux to below 600 kHz count rate per detector.

Phase information from both the WDM and TDM channels is processed by an FPGA‑based servo controller executing our prior‑assisted Bayesian estimation (Fig.~\ref{fig:setup}c). This system converts time‑stamped photon‑detection events into minimal‑variance phase estimates and applies corrective feedback via AOMs placed in the interferometer arms. AOM 2, positioned in the common path before demultiplexing, provides rapid correction for fiber‑induced noise tracked by the WDM channel. AOM 3, placed in the quantum path after demultiplexing, corrects the overall phase offset between the two interferometer arms. A fixed‑frequency AOM 1 ensures the 1548 nm references from both arms arrive at Charlie with identical optical frequencies, maintaining a stable interference fringe for the WDM stabilization system.

The core Bayesian estimator employs strong pre-characterization probes to measure the system's phase-diffusion coefficient $D$. It then models the diffusive phase noise as a Wiener process, wherein the phase increment over a measurement interval $\tau$ follows $\Delta\phi \sim \mathcal{N}(0, D\tau)$. This model implies a variance scaling $\sigma_T^2 = kD\tau$ over a duration $T = k\tau$, establishing the mathematical prior for optimal tracking. Subsequently, the estimator assesses the statistical consistency of each measurement innovation against this prior using a threshold multiplier $\kappa$.

\begin{figure*}[ht!]
\centering
\includegraphics[width=17cm]{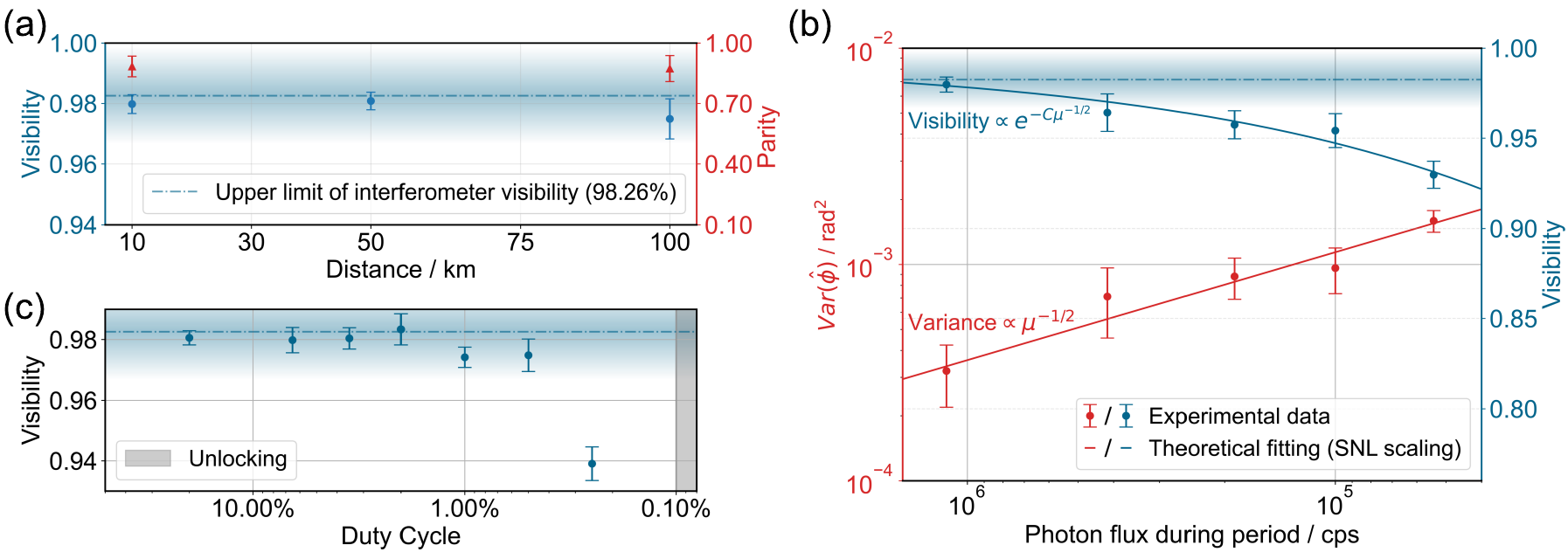}
\caption{\textbf{Experimental results.}
\textbf{(a)} Interferometric visibility (blue circles) and parity contrast (red triangles) versus fiber distance. The parity contrast is evaluated at the 10‑km and 100‑km links, while the visibility additionally includes a 50‑km intermediate point.
The horizontal blue dashed line indicates the maximum achievable interferometric visibility ($98.26\%$) of the system. Error bars represent $\pm1$ standard deviation from multiple experimental realizations.
\textbf{(b)} Residual phase variance (red circles) and interferometric visibility (blue circles) versus detected photon flux for the 10 km link. The residual variance follows the theoretical SNL prediction (red line), scaling as $\mu^{-1/2}$, confirming quantum‑optimal tracking.
\textbf{(c)} Interferometric visibility versus TDM duty cycle for the 10 km link, demonstrating robustness across operational parameters.}
\label{fig:exp_result}
\end{figure*}

{\it Results.---} We characterize phase diffusion between nodes by measuring diffusion coefficients $D_l$ for individual fiber spans and $D_r$ for total residual phase noise using continuous‑wave strong pre‑characterization probes, establishing Bayesian priors for subsequent experiments. Laboratory environmental fluctuations—predominantly acoustic and thermal coupling below 200 Hz—induce frequency deviations that accumulate over time. The measured phase variance scales as $\sigma_T^2 \propto T^{1.4}$ over extended durations $T$, deviating from the $\sigma_T^2 \propto T$ scaling of a pure Wiener process and indicating correlated (non‑white) noise (Supplemental Material~\cite{suppmat} S2.1).

As shown in Fig. \ref{fig:exp_result}(a), the system maintains phase stability with interferometric visibility exceeding $97.7\%$ over 10‑km fiber links (500 $\mu\text{s}$ experimental sequence, 32.5 $\mu\text{s}$ probe duration at 393 nm) and $97\%$ over 100‑km links (800 $\mu\text{s}$ sequence, 47.5 $\mu\text{s}$ probe duration) under realistic laboratory noise. This performance demonstrates SNL operation even under photon‑starved conditions. The high‑fidelity phase control enables deterministic ion–ion entanglement generation with a measured parity contrast exceeding $85\%$ at an excitation probability $\alpha = 5\%$, meeting stringent requirements for scalable quantum repeaters.

Fig.~\ref{fig:exp_result}(b) shows phase tracking across varying photon flux over 10 km. The residual variance follows the theoretical SNL prediction, scaling as $\mu^{-1/2}$, confirming quantum‑optimal tracking. Fig.~\ref{fig:exp_result}(c) demonstrates phase tracking across varying duty cycles with TDM stabilization over 10 km. SNL performance is maintained down to a duty cycle of $2\%$, with phase‑unlocking events occurring only below $0.1\%$ (where system noise dominates). This illustrates the framework’s robustness under different operational conditions and its ability to completely eliminate phase‑unlocking events that consistently occur with conventional MLE under identical noise conditions. 

Systematic limitations currently bound the maximum achievable visibility to $98.26\% \pm 1.58\%$, primarily due to electro-optic modulation distortion in the 854-nm excitation system and nonlinear noise in the frequency-conversion references. The observed visibilities of $97.98\% \pm 0.31\%$ (10\,km) and $97.49\% \pm 0.66\%$ (100\,km) fall slightly below this limit owing to residual phase-stabilization variance, imperfect temporal-mode matching, detector-efficiency nonuniformity, and the finite servo bandwidth imposed by the low-duty-cycle TDM sequence. Dedicated calibration measurements confirm that careful optimization can recover the system-limited upper bound (Supplemental Material~\cite{suppmat} S2.6).

{\it Discussion and Outlook.---} We demonstrate shot-noise limited phase stabilization between separate trapped-ion nodes, enabling a two-step excitation protocol for heralded, high-fidelity memory-memory entanglement generation that satisfies the fundamental requirement for scalable quantum repeaters~\cite{liu2026long}. The Bayesian estimation framework minimizes the required photon budget and duty cycle of reference probes—a critical feature for preserving fragile quantum states in large-scale networks. Furthermore, the framework exhibits resilience to non-ideal, correlated phase noise, underscoring its practical viability for real-world deployment.

Future enhancements could involve integrating machine-learning techniques~\cite{nolan2021machine} to adapt the phase-diffusion prior in real time, suppressing technical imperfections and extending robust operation to dynamic field environments. Beyond quantum networking, the framework’s photon efficiency could lengthen observational baselines in astronomical interferometry for faint sources~\cite{PhysRevLett.104.211103} and enhance phase tracking in precision metrology~\cite{udem2002optical} under low-signal conditions.

\section{Acknowledgments}

This work was supported by the Quantum Science and Technology-National Science and Technology Major Project (2024ZD0300202, 2021ZD0300802, 2023ZD0300100), 
the National Natural Science Foundation of China (Grant Nos. T2125010, 12374470
), the Chinese Academy of Sciences, the Key R\&D Plan of Shandong Province (Grant No. 2023CXPT105), Shandong provincial natural science foundation (Grant Nos. ZR2022LLZ006, ZR2022LLZ011). Q.Z. acknowledges support from the Taishan Scholar Program of Shandong Province and the XPLORER Prize from the New Cornerstone Science Foundation. M.Y.Z, F.X.C. and J.P.C. acknowledges support from the Young Expert Program of the Taishan Scholar Program of Shandong Province.

Guang-Cheng Liu, Chao-Hui Xue, Fa-Xi Chen contributed equally to this work.

\nocite{10.1115/1.3658902,PhysRevA.79.053843,Riley2008HandbookOF,PhysRevA.87.043833,sackett2000experimental,PhysRevA.105.033101,apolin2026recoil,saha2025high,slodivcka2012interferometric,nadlinger2022a,PhysRevLett.124.110501} 
\bibliographystyle{unsrt}
\bibliography{main}

\end{document}


\title{Supplemental Material: Bayesian Phase Stabilization at the Shot‑Noise Limit for Scalable Quantum Networks} %

\affiliation{Hefei National Research Center for Physical Sciences at the Microscale and School of Physical Sciences, University of Science and Technology of China, Hefei, Anhui 230026, China}
\affiliation{Hefei National Laboratory, University of Science and Technology of China, Hefei, China}
\affiliation{Jinan Institute of Quantum Technology and CAS Center for Excellence in Quantum Information and Quantum Physics, University of Science and Technology of China, Jinan 250101, China}
\affiliation{Shanghai Research Center for Quantum Science and CAS Center for Excellence in Quantum Information and Quantum Physics, University of Science and Technology of China, Shanghai, China}

\author{Guang-Cheng Liu}
\affiliation{Hefei National Research Center for Physical Sciences at the Microscale and School of Physical Sciences, University of Science and Technology of China, Hefei, Anhui 230026, China}
\affiliation{Hefei National Laboratory, University of Science and Technology of China, Hefei, China}
\affiliation{Shanghai Research Center for Quantum Science and CAS Center for Excellence in Quantum Information and Quantum Physics, University of Science and Technology of China, Shanghai, China}

\author{Chao-Hui Xue}
\affiliation{Hefei National Research Center for Physical Sciences at the Microscale and School of Physical Sciences, University of Science and Technology of China, Hefei, Anhui 230026, China}
\affiliation{Hefei National Laboratory, University of Science and Technology of China, Hefei, China}
\affiliation{Jinan Institute of Quantum Technology and CAS Center for Excellence in Quantum Information and Quantum Physics, University of Science and Technology of China, Jinan 250101, China}

\author{Fa-Xi Chen} %
\author{Ming-Yang Zheng}%
\affiliation{Hefei National Laboratory, University of Science and Technology of China, Hefei, China}
\affiliation{Jinan Institute of Quantum Technology and CAS Center for Excellence in Quantum Information and Quantum Physics, University of Science and Technology of China, Jinan 250101, China}

\author{Yi Yang}
\affiliation{Hefei National Research Center for Physical Sciences at the Microscale and School of Physical Sciences, University of Science and Technology of China, Hefei, Anhui 230026, China}
\affiliation{Hefei National Laboratory, University of Science and Technology of China, Hefei, China}
\affiliation{Jinan Institute of Quantum Technology and CAS Center for Excellence in Quantum Information and Quantum Physics, University of Science and Technology of China, Jinan 250101, China}

\author{Li-Bo Li}%
\affiliation{Jinan Institute of Quantum Technology and CAS Center for Excellence in Quantum Information and Quantum Physics, University of Science and Technology of China, Jinan 250101, China}

\author{Bin Wang} %
\affiliation{Jinan Institute of Quantum Technology and CAS Center for Excellence in Quantum Information and Quantum Physics, University of Science and Technology of China, Jinan 250101, China}

\author{Bo-Wen Yang}
\author{Hai-Feng Jiang}%
\author{Yong Wan}
\author{Ye Wang}
\affiliation{Hefei National Research Center for Physical Sciences at the Microscale and School of Physical Sciences, University of Science and Technology of China, Hefei, Anhui 230026, China}
\affiliation{Hefei National Laboratory, University of Science and Technology of China, Hefei, China}
\affiliation{Shanghai Research Center for Quantum Science and CAS Center for Excellence in Quantum Information and Quantum Physics, University of Science and Technology of China, Shanghai, China}

\author{Jiu-Peng Chen}
\affiliation{Hefei National Laboratory, University of Science and Technology of China, Hefei, China}
\affiliation{Jinan Institute of Quantum Technology and CAS Center for Excellence in Quantum Information and Quantum Physics, University of Science and Technology of China, Jinan 250101, China}

\author{Qiang Zhang}
\affiliation{Hefei National Research Center for Physical Sciences at the Microscale and School of Physical Sciences, University of Science and Technology of China, Hefei, Anhui 230026, China}
\affiliation{Hefei National Laboratory, University of Science and Technology of China, Hefei, China}
\affiliation{Jinan Institute of Quantum Technology and CAS Center for Excellence in Quantum Information and Quantum Physics, University of Science and Technology of China, Jinan 250101, China}
\affiliation{Shanghai Research Center for Quantum Science and CAS Center for Excellence in Quantum Information and Quantum Physics, University of Science and Technology of China, Shanghai, China}

\author{Jian-Wei Pan}
\affiliation{Hefei National Research Center for Physical Sciences at the Microscale and School of Physical Sciences, University of Science and Technology of China, Hefei, Anhui 230026, China}
\affiliation{Hefei National Laboratory, University of Science and Technology of China, Hefei, China}
\affiliation{Shanghai Research Center for Quantum Science and CAS Center for Excellence in Quantum Information and Quantum Physics, University of Science and Technology of China, Shanghai, China}

\maketitle

\tableofcontents

\section{Fisher information advantage for phase tracking}
\label{sec:fisher}


We develop a Fisher-information (FI) analysis for optical phase estimation and show that a prior-assisted recursive Bayesian protocol can approach the shot-noise limit (SNL), thereby systematically outperforming conventional maximum-likelihood estimation (MLE).

\subsection{Phase estimation model and Fisher information}
\label{subsec:measurement}

We model optical phase estimation in a Mach-Zehnder interferometer illuminated by weak coherent states at the single-photon level.
The mean photon counts detected $\lambda$ at the two output ports over integration time $\tau$ are governed by the interference equations
\begin{align}
\lambda_1(\phi) &= \frac{N}{2}\left(1 + V_0\cos\phi\right), \\
\lambda_2(\phi) &= \frac{N}{2}\left(1 - V_0\cos\phi\right),
\end{align}
where $N = \mu\tau$ represents the total mean photon number incident during measurement interval $\tau$, $\mu$ denotes the incident photon flux, and $V_0$ characterizes the maximum achievable interference visibility determined by system imperfections. 

The photon-counting events $n_1$ and $n_2$ are treated as statistically independent Poisson variables, giving the joint likelihood
\begin{equation}
\mathcal{L}(n_1,n_2 | \phi)  = \frac{[\lambda_1(\phi)]^{n_1} e^{-\lambda_1(\phi)}}{n_1!} \cdot \frac{[\lambda_2(\phi)]^{n_2} e^{-\lambda_2(\phi)}}{n_2!}.
\end{equation}

For any unbiased estimator of $\phi$, the estimation variance is bounded by the Cramér-Rao lower bound, $\sigma_{\phi}^2 \ge 1/I_F(\phi)$, where $I_F(\phi)$ is the FI extracted from the measurement statistics.
For our specific Poisson measurement model, the FI takes the canonical form
\begin{equation}
I_F(\phi) = \sum_{i=1}^2 \frac{1}{\lambda_i(\phi)} \left( \frac{\partial \lambda_i(\phi)}{\partial \phi} \right)^2.
\end{equation}
Computing the parametric derivatives of the mean photon count rates with respect to the phase parameter
\begin{align}
\frac{\partial \lambda_1(\phi)}{\partial \phi} &= -\frac{N V_0}{2} \sin\phi, \\
\frac{\partial \lambda_2(\phi)}{\partial \phi} &= \frac{N V_0}{2} \sin\phi.
\end{align}
Substitution gives the FI
\begin{align}
I_F(\phi) &= \left( \frac{N V_0}{2} \sin\phi \right)^2 \left[ \frac{1}{\lambda_1(\phi)} + \frac{1}{\lambda_2(\phi)} \right] \\
&= \frac{N V_0^2 \sin^2\phi}{2} \left[ \frac{1}{1+V_0\cos\phi} + \frac{1}{1-V_0\cos\phi} \right] \\
&= \frac{N V_0^2 \sin^2\phi}{1 - V_0^2 \cos^2\phi}.
\end{align}

The FI is maximized at the quadrature point $\phi=\pi/2$, giving

\begin{equation}
I_F(\pi/2) = N V_0^2 = \mu V_0^2 \tau.
\end{equation}

The variance of any unbiased phase estimator therefore satisfies

\begin{equation}
\sigma^2_{\mathrm{static}} \ge \frac{1}{\mu V_0^2 \tau},
\end{equation}
which rigorously defines the \textbf{SNL} for static phase measurement and manifests the characteristic shot-noise scaling $\sigma_\phi \propto 1/\sqrt{\mu}$ inherent in coherent state interferometry.

\subsection{Phase diffusion and the precision-delay trade-off}
\label{subsec:diffusion}

In the experiment, environmental perturbations and system instabilities cause the optical phase to diffuse. We model this evolution as a Wiener process with diffusion coefficient $D$.

Under these stochastic dynamics, the variance of the phase drift grows linearly with time: $\text{Var}(\delta\phi(t)) = D t$. For a diffusing phase measured over a finite integration time $\tau$, temporal averaging reduces the effective interference visibility. For a Wiener process, the phase deviation follows a $\delta\phi \sim \mathcal{N}(0, D\tau)$, yielding the statistical expectation
\begin{equation}
\langle \cos(\delta\phi) \rangle = \Re\left[ \langle e^{i\delta\phi} \rangle \right] = e^{-D\tau/2}.
\end{equation}

Phase averaging over the integration window therefore gives the effective visibility

\begin{equation}
V(\tau) = V_0 e^{-D\tau/2}.
\end{equation}

\subsection{Conventional maximum‑likelihood estimation phase tracking limits}
\label{subsec:conventional}

We first consider the phase-stabilization update cycle, in which the feedback applied at step $k+1$ is based on the measurement obtained at step $k$. The phase evolves as $\phi_{k+1} = \phi_{k}-\hat{\phi}_{\text{meas},k}+\delta\phi_k$, where $\delta \phi_k \sim \mathcal{N}(0,D\tau)$ is the process noise. Under phase diffusion, substituting $V(\tau)$ into the static expression gives the average FI extracted from a measurement of duration $\tau$.
\begin{equation}
\bar{I}_{\mathrm{F}}(\tau) = \mu [V(\tau)]^2 \tau = \mu V_0^2 \tau e^{-D\tau}.
\end{equation}

For conventional MLE phase-tracking protocols, the total phase error reflects a precision-delay trade-off arising from two statistically independent contributions
\begin{equation}
    \text{Var}(\Phi) = \sigma^2_{\text{meas}}+\sigma^2_{\text{diffusion}}.
\end{equation}

\begin{enumerate}
\item \textbf{Measurement noise:} 

Bounded by the Cramér-Rao bound: $\sigma^2_{\mathrm{meas}} \gtrsim 1/\bar{I}_{\mathrm{F}}(\tau) = (\mu V_0^2 \tau e^{-D\tau})^{-1}$;
\item \textbf{Delay-induced diffusion:} 

Variance accumulated during signal acquisition: $\sigma^2_{\mathrm{diffusion}} = D\tau$.
\end{enumerate}
The combined lower bound for conventional phase tracking therefore satisfies
\begin{equation}
\sigma^2_{\mathrm{conv}}(\tau) \gtrsim \frac{1}{\mu V_0^2 \tau e^{-D\tau}} + D\tau.
\end{equation}

Minimizing this expression reveals the \textbf{precision-delay trade-off:} increasing $\tau$ reduces shot noise but increases diffusion during signal acquisition.

\subsection{Bayesian phase estimation for phase tracking with Fisher information advantage}
\label{subsec:bayesian}

Our protocol uses a prior-assisted recursive Bayesian estimator to relax the precision-delay trade-off in phase tracking. The estimator represents the phase state by a Gaussian posterior $p(\phi_k | \mathcal{D}_{1:k})$ with mean $\hat{\phi}_k$ and variance $\sigma_k^2$, where $\mathcal{D}_{1:k}$ denotes the measurement history up to step $k$. Each recursive update has two steps that combine the prior with the new photon-count data.
\begin{enumerate}
\item \textbf{Prediction Step:} 

Wiener-process phase diffusion with coefficient $D$ propagates the posterior variance as
\begin{equation}
\sigma^2_{\mathrm{pred}} = \sigma_k^2 + D\tau.
\end{equation}

This step accounts for phase uncertainty accumulated between successive measurements through the additive term $D\tau$.

\item \textbf{Update Step:} 

Update step: New photon-count data $n = (n_1, n_2)$ update the posterior through Bayes' theorem:

\begin{equation}
p(\phi_{k+1} | \mathcal{D}_{1:k+1}) \propto \mathcal{L}(\mathbf{n}_{k+1} | \phi_{k+1}) \cdot \mathcal{N}(\phi_{k+1}; \hat{\phi}_k, \sigma^2_{\mathrm{pred}}).
\end{equation}

where $\mathcal{L}(n_{k+1}|\phi_{k+1})$ is the Poisson likelihood and $\mathcal{N}(\phi_{k+1}; \hat{\phi}_k, \sigma^2_\text{pred})$ is the predicted prior incorporating phase diffusion.

When the photon flux is sufficient for the central-limit approximation, the Poisson likelihood 
$\mathcal{L}(\mathbf{n} | \phi)$ can be approximated by a Gaussian distribution with variance $I_{\mathrm{eff}}^{-1}$, yielding the efficient variance update
\begin{equation}
\sigma_{k+1}^{-2} = \left( \sigma_k^2 + D\tau \right)^{-1} + I_{\mathrm{eff}}. \label{eq:update}
\end{equation}
\end{enumerate}

Corresponding FI for steady state is
\begin{equation}
    I_{F}^{\text{eff}} = I^{\text{prior}}_F+ I^{\text{meas}}_F = \frac{1}{D\tau +\sigma^2_\infty}+\Gamma \tau e^{-D\tau},
\end{equation}
where $\Gamma = \mu V_0^2$. 
To incorporate prior information in the photon-starved regime, we apply a nonlinear innovation filter to the phase deviation, ($\delta\hat{\varphi}_k \equiv \hat{\phi}_{\text{meas},k}-\phi_0$)
\begin{equation}
f(\delta\hat{\varphi}_k) =
\begin{cases}
\delta\hat{\varphi}_k & |\delta\hat{\varphi}_k| \leq \kappa\sigma_{\text{prior}} \\
    \operatorname{sign}(\delta\hat{\varphi}_k) \cdot \left[\kappa\sigma_{\text{prior}} + \Delta \exp\left(-\dfrac{\Delta}{\sigma_{\text{prior}}}\right)\right] & |\delta\hat{\varphi}_k| > \kappa\sigma_{\text{prior}}.
\end{cases}
\label{eq:filter}
\end{equation}

Here, $\kappa$ is the threshold multiplier that defines the statistically plausible range $\kappa \sigma_{\rm prior}$, whereas $\Delta = |\delta\hat{\varphi}_k| - \kappa\sigma_{\text{prior}}$ quantifies the excess deviation beyond this threshold. The filter preserves innovations within the expected statistical envelope while exponentially suppressing outliers beyond the threshold. A detailed discussion of $\kappa$ follows in the next section.

The filter constrains the accepted innovations to the scale $\kappa\sigma_{\rm prior}$. After the $k\rightarrow k+1$ prediction step, we set $\sigma^2_{\text{prior}}= D\tau$. The effective measurement information after soft-filter mapping is
\begin{equation}
    I^{\text{meas},\text{out}}_{F} = \eta(\kappa)I_{F}^{\text{meas},\text{in}}.
\end{equation}

The effective FI from the prior and the measurement is
\begin{equation}
    I_{\text{eff}} = I_{F}^{\text{prior}}+I_{F}^{\text{meas}} = \frac{1}{D\tau+\sigma^2_\infty}+ \eta(\kappa)\Gamma \tau e^{-D\tau}.
\end{equation}

The recursive estimator reaches steady state when the variance becomes stationary, $\sigma_{k+1}^2=\sigma_k^2=\sigma_{\infty}^2$.
Substituting this stationary condition into  Eq.~\eqref{eq:update} yields
\begin{equation}
\sigma_{\infty}^{-2} = \left( \sigma_{\infty}^2 + D\tau \right)^{-1} + \Gamma\tau e^{-D\tau}\eta(\kappa). \label{eq:steady_state}
\end{equation}


Under the condition $D\tau \ll \sigma_{\infty}^2$, retaining the leading-order terms yields
\begin{equation}
D \approx \Gamma \eta(\kappa)\sigma_{\infty}^4 \quad \Rightarrow \quad \sigma_{\infty}^2 \approx (\eta(\kappa))^{-1/2}\sqrt{\frac{D}{\Gamma}} = \left(\frac{ D}{\mu V_0^2 \eta(\kappa)}\right)^{1/2}. \label{eq:sql_tracking}
\end{equation}

This expression gives the SNL scaling for tracking a diffusing phase under the stated model assumptions. The same precision bound can also be derived using the Riccati equation approach~\cite{10.1115/1.3658902,PhysRevA.79.053843}, which yields the scaling relation $\sigma^2_{\infty} \propto 1/\sqrt{\mu}$ that improves over the conventional static phase-estimation scaling under the stated model assumptions $\sigma^2_{\text{conv}}\propto 1/\mu$.

When the measurement information is weak ($D\tau \ll \sigma^2_{\infty}$), the total estimation variance for the prior-assisted protocol, incorporating both the tracking limit and the diffusion during measurement, is given by
\begin{equation}
\sigma^2_{\mathrm{Bay}}(\tau) \gtrsim \frac{1}{I^\mathrm{prior}_{\text{eff}}+\bar{I}_F}+D\tau=\frac{1}{\frac{1}{D\tau + (\eta(\kappa))^{-1/2} \sqrt{D/\Gamma}}+\eta(\kappa)\mu V_0^2\tau e^{-D\tau}}+ D\tau. \label{eq:total_variance}
\end{equation}

The key advantage is that the tracking term is independent of the measurement delay $\tau$, allowing short integration times to reduce diffusion without the usual shot-noise penalty. This delay-independent scaling relaxes the precision-delay trade-off that limits conventional MLE approaches. 

\subsection{Bayesian phase tracking performance simulation}
\label{subsec:simulation}

To validate the estimator robustness, we performed numerical simulations of phase diffusion and phase-tracking performance. The simulations reproduce the relevant experimental dynamics via a discrete random walk with diffusion constant $D$: each phase increment $\delta \phi_k$ is drawn from $\mathcal{N}(0, D\tau)$, consistent with Wiener-process statistics.

At each measurement interval $\tau$, we simulated the detection process by drawing photon counts $n_1$ and $n_2$ at the two interferometer output ports from Poisson distributions
\begin{equation}
n_{1,2} \sim \mathrm{Poisson}\left(\lambda_{1,2}(\phi)\right), \quad \text{with} \quad \lambda_{1,2}(\phi) = \frac{N}{2}\left[1 \pm V_0\cos\phi\right],
\end{equation}
where $N = \mu\tau$ is the mean photon number per measurement. This procedure includes shot-noise and tests estimator performance under photon-starved conditions.

\begin{figure}[ht!]
\centering
\includegraphics[width=\linewidth]{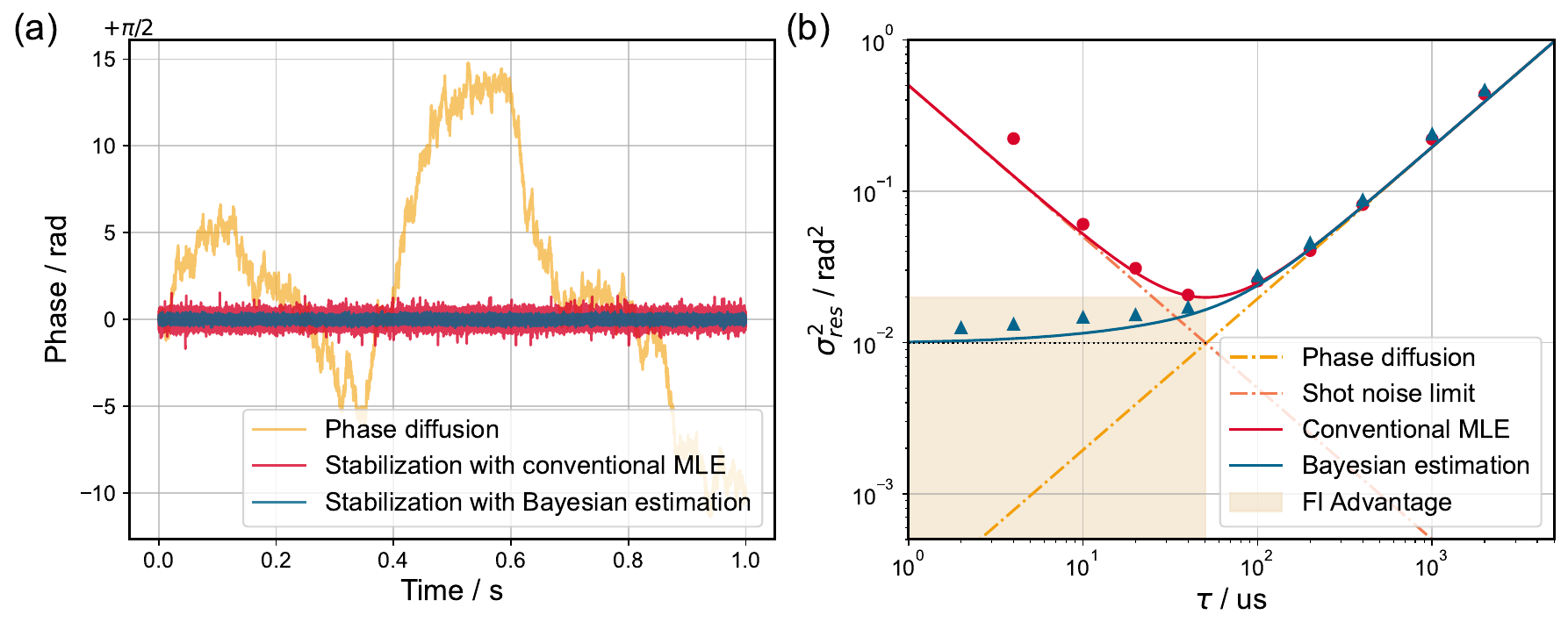}
\caption{\textbf{Numerical simulation of phase-stabilization  performance.} (a) Time series of phase evolution at 10 $\mu s$ integration time. The yellow curve represents uncontrolled free phase diffusion. The red and blue curve show phase stabilization using conventional MLE and the prior-assisted recursive  Bayesian estimator, respectively.
(b) Phase stabilization performance comparison. Scatter points and solid lines show the simulated performance of the conventional MLE (red solid, red circles) and the Bayesian estimation (blue solid, blue triangles). The yellow dot-dashed line indicates the theoretical variance of free phase diffusion and the pink dashed line marks the photon shot-noise limit. The shaded FI-advantage region highlights the regime in which prior information improves performance.
}
\label{fig:Fig_simulation_timeinfo}
\end{figure}

We simulated photon-starved operation with ($\mu=2$ MHz, $D=2\times10^{-4}\  \text{rad}^2/\mu s,\kappa=1$)  to stabilize the phase at $\pi/2$. The Bayesian estimator suppresses phase diffusion more effectively in the time domain (Fig.~\ref{fig:Fig_simulation_timeinfo}(a)) and yields lower estimation variance across the simulated integration times (Fig.~\ref{fig:Fig_simulation_timeinfo}(b)). At short integration times, the Bayesian estimator enters the FI-advantage region shaded in Fig.\,1 of main text, defined as the regime below the shot-noise--diffusion crossover time where the residual phase variance remains lower than that of conventional MLE. This advantage arises because the stabilized prior from previous updates compensates for reduced photon-counting information, changing the scaling from the shot-noise-limited $1/\mu$ behavior toward the dynamic-tracking $1/\sqrt{\mu}$ scaling. These results indicate that prior information is most beneficial in the FI-advantage regime, where photon counts are sparse.

\begin{figure}[ht!]
\centering
\includegraphics[width=0.6\linewidth]{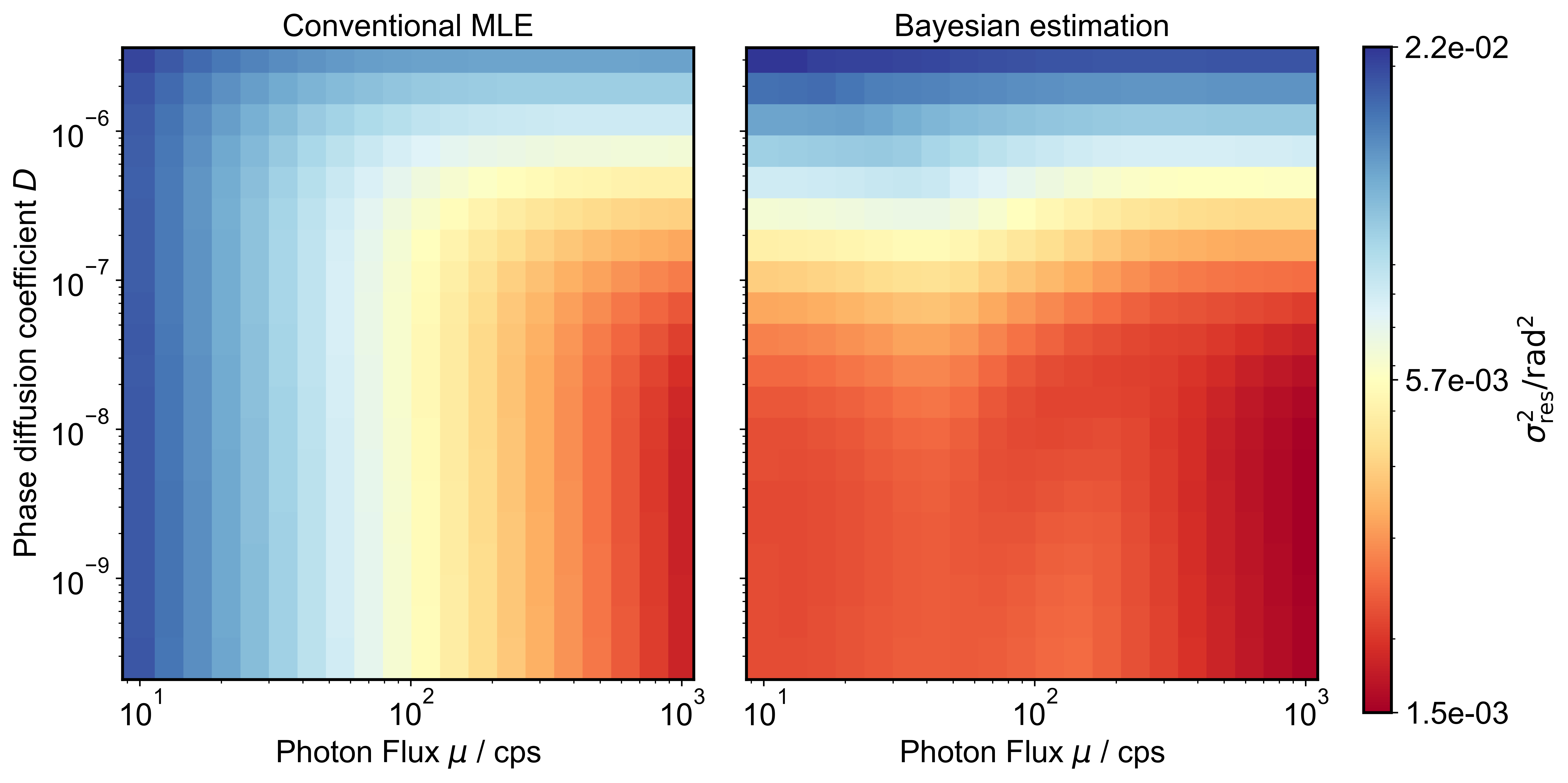}
\caption{\textbf{Performance comparison of phase-tracking protocols.} Residual phase variance (rad$^2$, color scale) as a function of photon flux $\mu$ and diffusion  coefficient $D$ for (left) conventional MLE and (right) prior-assisted estimation. The measurement window duration is fixed at 100\,$\mu$s. The prior-assisted approach maintains robust performance across photon-starved regimes where conventional methods degrade due to measurement outliers.
}
\label{fig:Fig_simulation_comparison}
\end{figure}

After each simulated measurement, we applied phase correction using both conventional MLE and the prior-assisted Bayesian protocol. We quantified feedback performance using the residual phase variance $\sigma^2_{\mathrm{res}} = \langle(\phi_{\mathrm{true}} - \hat{\phi})^2\rangle$, averaged over multiple independent trajectories.
Fig.~\ref{fig:Fig_simulation_comparison} compares conventional MLE with the prior-assisted estimator. The simulations identify the regimes in which the prior-assisted estimator reduces the residual phase variance relative to conventional MLE.

Conventional MLE degrades at low photon flux, where shot-noise generates measurement outliers that destabilize the feedback loop. By contrast, the Bayesian estimator remains stable across the simulated flux and diffusion ranges. The nonlinear innovation filter provides adaptive suppression of statistical outliers while preserving responsiveness to genuine phase variations. This advantage is most pronounced under photon-starved conditions, where the filter prevents spurious large corrections that would otherwise amplify phase fluctuations. 

\begin{figure}[ht!]
\centering
\includegraphics[width=0.8\linewidth]{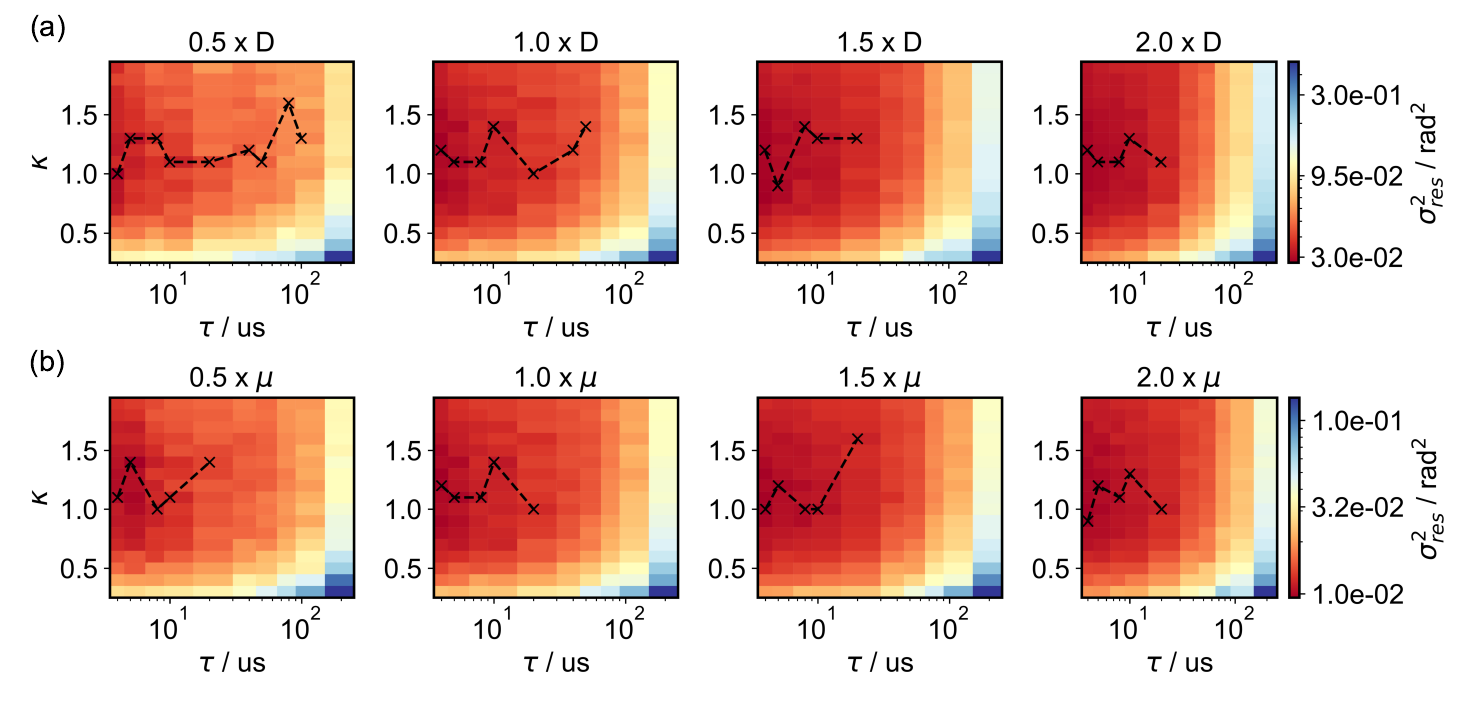}
\caption{\textbf{Determining the optimal range for the threshold multiplier $\kappa$.} Residual phase variance (rad$^2$, color scale) as a function of the threshold multiplier $\kappa$ and the measurement window duration $\tau$ for two constrained scenarios: (a) with a fixed diffusion coefficient $D$ and (b) with a fixed photon flux $\mu$, where each subplot shares a common color scale. The optimal $\kappa$ value for each condition is traced by the dashed black line. The simulations identify $1\lesssim\kappa\lesssim1.5$ as a suitable range in the FI-advantage region.}

\label{fig:Fig_simulation_kappa}
\end{figure}

We perform steady-state simulations to identify the suitable values of the threshold multiplier $\kappa$ for different combinations of phase diffusion coefficient $D$, photon flux $\mu$, and measurement window duration $\tau$. Acting as a consistency check, the threshold multiplier $\kappa$ bridges prior assumptions and experimental phase statistics. It defines the expected envelope of phase fluctuations. Our numerical analysis (Fig.~\ref{fig:Fig_simulation_kappa}) identifies ($1\lesssim\kappa\lesssim1.5$) as its effective operational range. In practice, this range provides a robustness buffer against deviations in the experimentally determined parameters and helps maintain reliable performance amid statistical variation.

The simulations indicate that the Bayesian estimator extends stable operation into the FI-advantage regime and maintains residual phase variance under conditions in which conventional MLE becomes unstable. This stability supports quantum-network protocols that require sustained optical phase coherence.

\clearpage

\section{SNL phase tracking enables high-fidelity entanglement generation between trapped ions}
\label{sec:parity}

We tested the phase-estimation protocol in a dual-band stabilization architecture that maintains inter-node phase coherence between two fiber-linked trapped-ion quantum nodes. Performance was assessed by measuring interferometric visibility and parity contrast during single-photon-interference entanglement generation. These observables provide complementary measures of phase stability and entanglement fidelity.

\subsection{Phase noise analysis and prior-assisted parameter}

\begin{figure}[ht!]
\centering
\includegraphics[width=\linewidth]{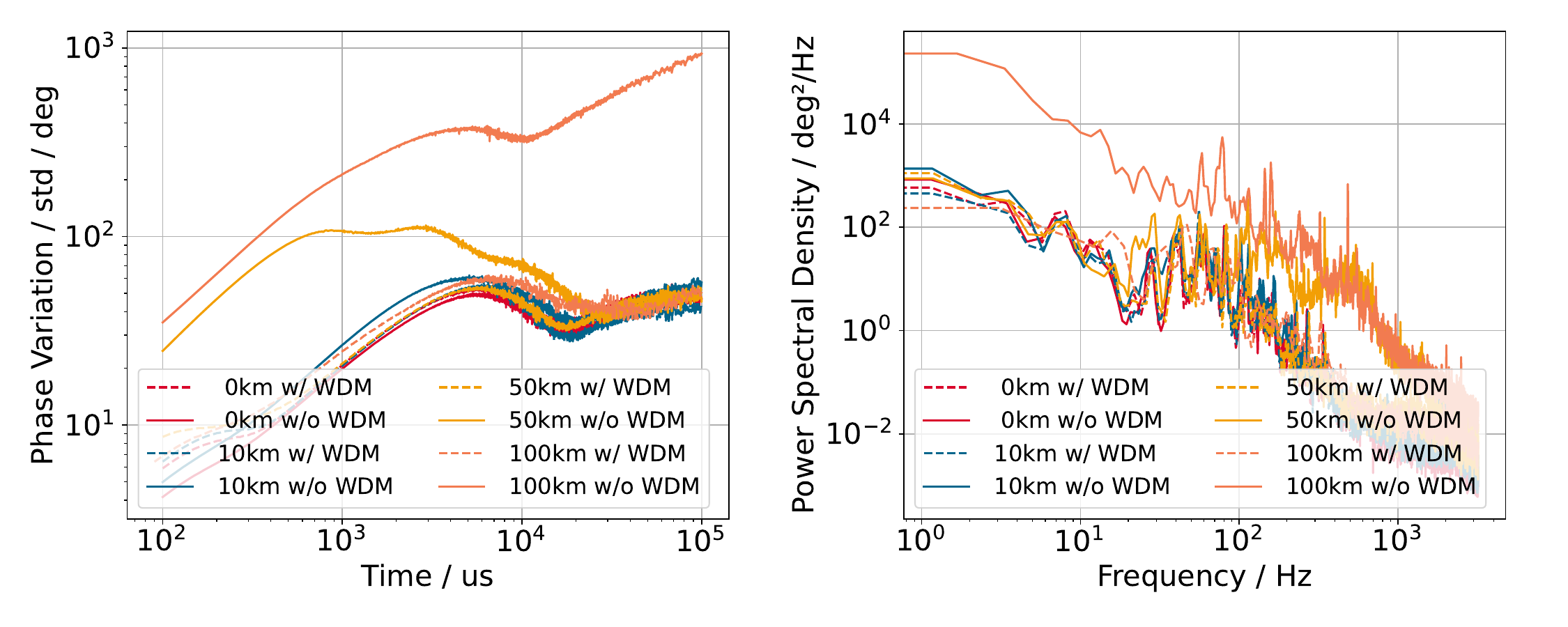}
\caption{\textbf{Phase evolution and noise power spectral density.} The measured phase evolution over time (left) and its corresponding PSD (right) are shown for transmission distances of 0 km, 10 km, 50 km, and 100 km. The solid and dashed lines distinguish the results obtained with the wavelength-division multiplexed (WDM) stabilization system disabled (w/o WDM) and enabled (w/ WDM), respectively.}
\label{fig:Fig_phase_1}
\end{figure}

The development of high-precision phase locking is limited by various phase fluctuations. To characterize these fluctuations, we used heterodyne detection, which avoids phase wrapping in homodyne measurements. The measured phase drift had a standard deviation exceeding $10^\circ$ over a 1 ms interval. Such short-term drift poses a major challenge for high-precision phase stabilization. The temporal phase evolution, derived from I-Q demodulation of the heterodyne signal, and its corresponding power spectral density are presented in Fig.~\ref{fig:Fig_phase_1}. 

\begin{figure}[ht!]
\centering
\includegraphics[width=0.5\linewidth]{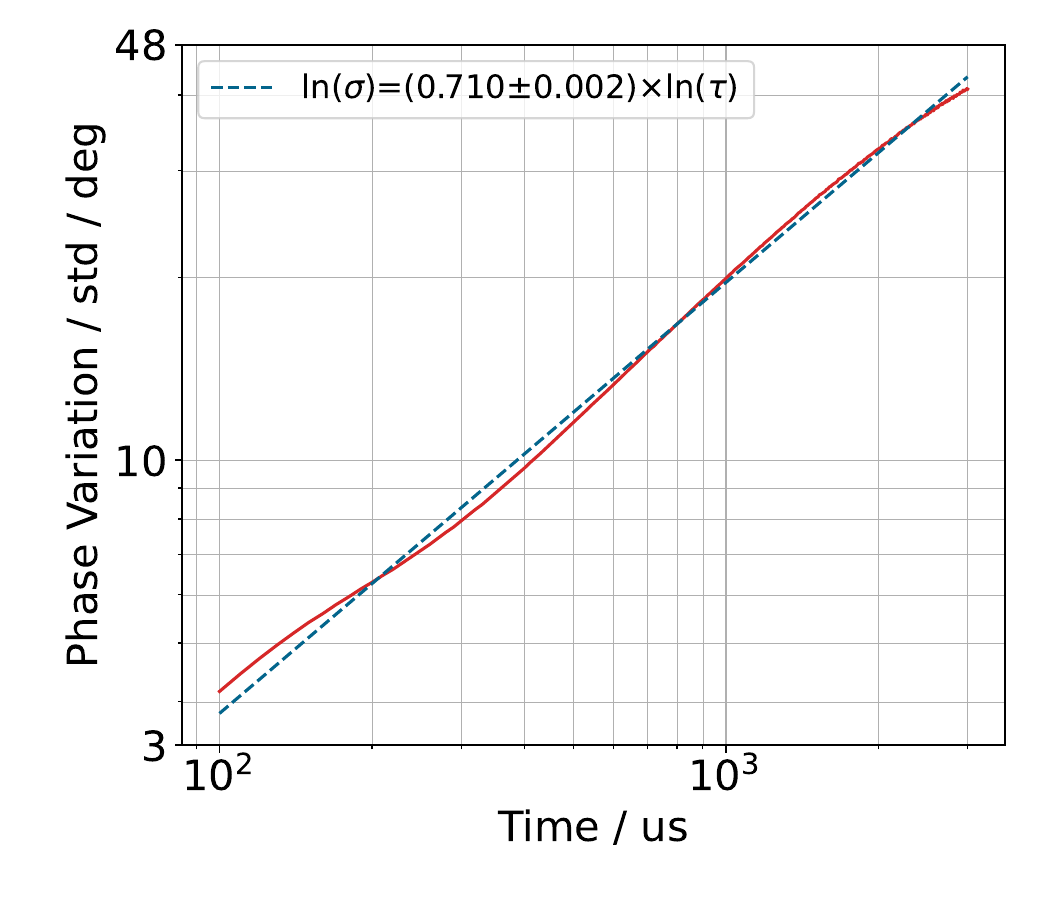}
\caption{\textbf{Phase evolution at 0 km (baseline, no additional fiber link)}. The data presented on a log-log scale exhibits a linear fit with a slope of 0.710 ± 0.002, indicating a power-law exponent of $\sigma\propto\tau^{0.710}$. The intercept of this fit yields a phase diffusion coefficient of $D=(0.15 ^{\circ}/us^{0.710})^2$, which is used directly to determine the prior parameters for the filter.}
\label{fig:Fig_phase_2}
\end{figure}

The phase variation $\sigma(\tau)$ quantifies the fluctuation magnitude via the sample standard deviation of adjacent differences in non-overlapping averaged phase sequences, given by $\sigma(\tau) = \text{Std}(\Delta\bar{\varphi}(\tau))$. Fig.~\ref{fig:Fig_phase_2} reveals a power-law scaling $\sigma(\tau) \propto \tau^{0.710}$, distinctly exceeding the $\tau^{0.5}$ signature of random-walk noise. This exponent, intermediate between diffusion ($0.5$) and linear drift ($1.0$), demonstrates long-range correlations in the phase dynamics inconsistent with a simple Wiener process model.

The temporal evolution of phase variation is described by the power-law model:
\begin{equation}
\sigma(\tau)=\sqrt{D}\tau^{\beta}
\end{equation}
which represents power-law noise in the frequency domain~\cite{Riley2008HandbookOF}. Here, $\beta$ is the scaling exponent that characterizes the noise, and $\sqrt{D}$ is a coefficient with units of $^\circ/\mu s^{\beta}$. This relationship manifests as a linear increase on a $\ln{\left(\sigma\right)}-\ln{\left(\tau\right)}$ scale with slope $\beta$ and intercept $\frac{1}{2}\ln{\left(D\right)}$. The value of $\beta$ distinguishes between different noise types: white phase noise corresponds to $\beta=-0.5$, yielding units of [$^\circ/\mu s^{-0.5}$] (equivalent to a flat amplitude spectral density in $\left[\mathrm{rad}/\sqrt{\mathrm{Hz}}\right]$); an ideal Wiener process corresponds to $\beta=0.5$, where $\sqrt D$ reduces to the conventional diffusion coefficient $\sqrt{D_{\mathrm{Wiener}}}$ with units of [$^\circ/\sqrt{\mu s}$]. For our experimental conditions, the fitted exponents range from $\beta\approx0.61$ to $0.85$, indicating the presence of correlated (colored) noise components arising from multiple environmental sources. These estimates of $\beta$ and $\sqrt{D}$ are used to obtain representative prior-assisted parameters, providing the Bayesian estimator with typical values.

The fit provides a quantitative calibration: its intercept yields the diffusion coefficient $D$, which in turn determines the prior parameter $\sigma$ for the filter at a $\tau_{0}=50\, \mu s$ integration time, $\sigma=\sqrt{D}\tau_0^{\beta}$. We applied this calibration protocol uniformly across all experimental configurations. The derived prior parameters $\sigma$ for 10 km, 50 km, and 100 km fiber lengths, which enable optimal filtering under different fiber noise characteristics, are compiled in Table~\ref{tab:prior_param}.

Analysis of the phase-noise power spectral density (PSD) revealed multiple discrete, high-amplitude spectral peaks. We attribute these peaks to acoustic and mechanical disturbances in the laboratory environment, primarily the secondary air-conditioning system (at tens of Hz) and building vibrations. (ranging from a few Hz to hundreds of Hz). The observed scaling exponent greater than 0.5 is thus interpreted as the result of characteristic-frequency noise components elevating the overall scaling above the random-walk baseline.

In the single-frequency noise model, $\varphi(t)=A \sin(2\pi f t)$ represents a monochromatic perturbation with amplitude $A$ and frequency $f$. Its effect on the measured phase variation ($\text{Std}{(\Delta\bar{\varphi}(\tau))}$) over an interval $\tau$ is characterized by the envelope response $\sigma(\tau) = A \sin^2(\pi f \tau)/(\pi f \tau)$. The local scaling exponent, defined as $\beta(\tau) = d(\ln\sigma)/d(\ln\tau) = 2\pi f \tau \cot(\pi f \tau) - 1$, reveals the underlying scaling behavior: for $f\tau \ll 1$, $\beta(\tau) \to 1$, giving drift-like scaling ($\mathrm{std} \propto \tau$). This mechanism explains how narrowband noise can shift the aggregate exponent from the diffusive baseline of 0.5 toward 1. Nevertheless, because $\sigma(\tau) \to 0$ as $\pi f \tau \to 0$, the relative contribution of the narrowband component becomes negligible at sufficiently short timescales. Thus, although narrowband noise elevates the effective exponent toward 1, its diminishing amplitude allows the underlying diffusive noise to dominate in the short-$\tau$ limit, restoring the aggregate scaling to $\beta \to 0.5$.

The experimental results in Table~\ref{tab:prior_param} show that the scaling exponent $\beta$ derived from measured phase noise increases with fiber length, tending towards a value of 1. This trend suggests that longer fibers act as larger antennas for low-frequency, environmentally correlated noise, such as temperature fluctuations and vibrations, causing the phase evolution to more closely resemble linear drift. The statistical uncertainties for $\beta$ in Table~\ref{tab:prior_param} are derived from the spread of estimates across multiple temporal segments of a continuous data stream, rather than from linear regression fitting errors. These values serve as a valid lower bound for consistent comparison across configurations, though they may slightly underestimate the true long-term variability due to the finite acquisition duration relative to slow environmental drifts. The distribution of $\beta$ itself reflects the stability of laboratory noise sources over the measurement window. Separately, the calibration parameter $D$ listed in Table~\ref{tab:prior_param} is obtained from a single representative dataset and should be interpreted as a typical value.

\begin{table}[h]
\centering
\caption{\textbf{Prior-assisted parameters $\sigma$, diffusion coefficients $D$ and scaling exponent $\beta$.} This table presents the prior parameters ($\sigma$), diffusion coefficients ($D$) and scaling exponent ($\beta$), where $_l$ corresponds to various fiber lengths and $_r$ to residual phase fluctuations with WDM stabilization. All parameters are derived for a $50\,\mu s$ integration time. The listed $D$ values are selected as typical calibrations from the experiments.
}
\begin{tabular}{ccccc}
\toprule
\textbf{parameter/coefficient} & \textbf{0 km} & \textbf{10 km } & \textbf{50 km} & \textbf{100 km} \\ 
\midrule
$\sigma_l$ & $2.3^{\circ}$  & $2.4^{\circ}$ & $15.0^{\circ}$ & $19.5^{\circ}$ \\
$\sigma_r$ & $2.1^{\circ}$ & $2.3^{\circ}$ & $3.3^{\circ}$ & $3.1^{\circ}$\\
$\sqrt{D_l}$  &  $0.15^{\circ}/\mu s^{0.71{0}}$   &  $0.09^{\circ}/\mu s^{0.82{0}}$  &  $0.64^{\circ}/\mu s^{0.8{07}}$  &  $0.70^{\circ}/\mu s^{0.85{1}}$ \\
$\sqrt{D_r}$  &  $0.11^{\circ}/\mu s^{0.76{4}}$  &  $0.14^{\circ}/\mu s^{0.7{19}}$  &  $0.29^{\circ}/\mu s^{0.6{16}}$  &  $0.22^{\circ}/\mu s^{0.68{4}}$\\

$\beta_l$  & $0.711 \pm 0.003$ & $0.803 \pm 0.035$ & $0.825 \pm 0.042$ & $0.850 \pm 0.010$ \\

$\beta_r$  & $0.759 \pm 0.004$ & $0.717 \pm 0.005$ & $0.612 \pm 0.006$ & $0.689 \pm 0.008$ \\
\bottomrule
\end{tabular}

\label{tab:prior_param}
\end{table}

\subsection{Generalized Bayesian tracking limit for power-law phase diffusion}
\label{subsec:power_law_bayesian_limit}

The Wiener-process treatment in the main text represents the ideal random-walk limit of phase diffusion. In practice, environmental perturbations and system instabilities introduce correlated phase noise, which we capture with a power-law generalization of the accumulated phase fluctuation,
\begin{equation}
\sigma_\phi(\tau)=\sqrt{D_\beta}\tau^\beta.
\end{equation}
Equivalently, the phase-increment variance is
\begin{equation}
Q_\beta(\tau)\equiv \mathrm{Var}[\delta\phi(\tau)]
=D_\beta\tau^{2\beta}.
\end{equation}
Here, $\beta$ is the scaling exponent and $D_\beta$ is an empirically calibrated phase-noise scale coefficient whose units depend on $\beta$. The Wiener limit corresponds to $\beta=0.5$, where $D_\beta$ reduces to the ideal diffusion coefficient $D_{\mathrm{Wiener}}$ and $Q_\beta(\tau)=D_{\mathrm{Wiener}}\tau$.

Over the feedback timescale, we approximate the short-time phase increment as Gaussian
\begin{equation}
\delta\phi(\tau)\sim \mathcal{N}(0,Q_\beta),
\end{equation}
Temporal averaging then reduces the interference visibility to
\begin{equation}
V(\tau)
=
V_0\exp\left[-\frac{Q_\beta(\tau)}{2}\right]
=
V_0\exp\left[-\frac{D_\beta\tau^{2\beta}}{2}\right].
\end{equation}
The average FI collected during an integration time $\tau$ is therefore
\begin{equation}
\bar I_F(\tau)
=
\Gamma\tau\exp[-D_\beta\tau^{2\beta}],
\quad
\Gamma=\mu V_0^2 .
\end{equation}
For the recursive Bayesian estimator, the prediction step propagates the posterior variance as
\begin{equation}
\sigma_{\rm pred}^2
=
\sigma_k^2+D_\beta\tau^{2\beta}.
\end{equation}
Within the Gaussian-information approximation, the posterior after one measurement update is
\begin{equation}
\sigma_{k+1}^{-2}
=
\left(\sigma_k^2+D_\beta\tau^{2\beta}\right)^{-1}
+
\eta(\kappa)\Gamma\tau e^{-D_\beta\tau^{2\beta}},
\end{equation}
where $\eta(\kappa)$ accounts for the information retained after nonlinear innovation filtering. At steady state, $\sigma_{k+1}^2=\sigma_k^2=\sigma_\infty^2$, giving
\begin{equation}
\sigma_\infty^{-2}
=
\left(\sigma_\infty^2+D_\beta\tau^{2\beta}\right)^{-1}
+
\eta(\kappa)\Gamma\tau e^{-D_\beta\tau^{2\beta}} .
\end{equation}
Solving this quadratic equation for the positive variance gives the generalized Bayesian tracking limit
\begin{equation}
\sigma_{\infty,\rm Bay}^2(\tau)
=
\frac{1}{2}
\left[
\sqrt{
D_\beta^2\tau^{4\beta}
+
\frac{4D_\beta}{\eta(\kappa)\Gamma}
\tau^{2\beta-1}
e^{D_\beta\tau^{2\beta}}
}
-
D_\beta\tau^{2\beta}
\right].
\end{equation}
When the measurement information is weak, $D_\beta\tau^{2\beta}\ll\sigma_\infty^2$ and $D_\beta\tau^{2\beta}\ll1$, this expression reduces to
\begin{equation}
\sigma_{\infty,\rm Bay}^2
\approx
\left(
\frac{D_\beta}{\eta(\kappa)\mu V_0^2}
\right)^{1/2}
\tau^{\beta-1/2}.
\label{Eq:general_bay}
\end{equation}
For $\beta=0.5$, the dependence on $\tau$ vanishes and the result becomes
\begin{equation}
\sigma_{\infty,\rm Bay}^2
\approx
\left(
\frac{D_{\mathrm{Wiener}}}{\eta(\kappa)\mu V_0^2}
\right)^{1/2},
\end{equation}
which is the Wiener-process Bayesian tracking limit used in the simplified analytical treatment. Thus, the power-law exponent $\beta$ modifies only the integration-time dependence of the Bayesian tracking limit, while the photon-flux scaling of the steady-state variance remains $\sigma_{\infty,\rm Bay}^2\propto \mu^{-1/2}$ for any $\beta$ within this approximation.

\subsection{Dual-band stabilization architecture for inter-node phase coherence}

The single-photon interference scheme for entanglement generation achieves high efficiency but requires maintaining inter-node phase coherence against environmental perturbations. We overcome this challenge through a hierarchical dual-band stabilization architecture implementing our prior-assisted phase estimation protocol to utilize scarce probe photons. This approach targets distinct noise sources on different timescales while maintaining phase correlation between reference probes and quantum signals through a common-path design.

\begin{figure}[ht!] 
\centering
\includegraphics[width=1\linewidth,trim=0 10 0 0, 
    clip]{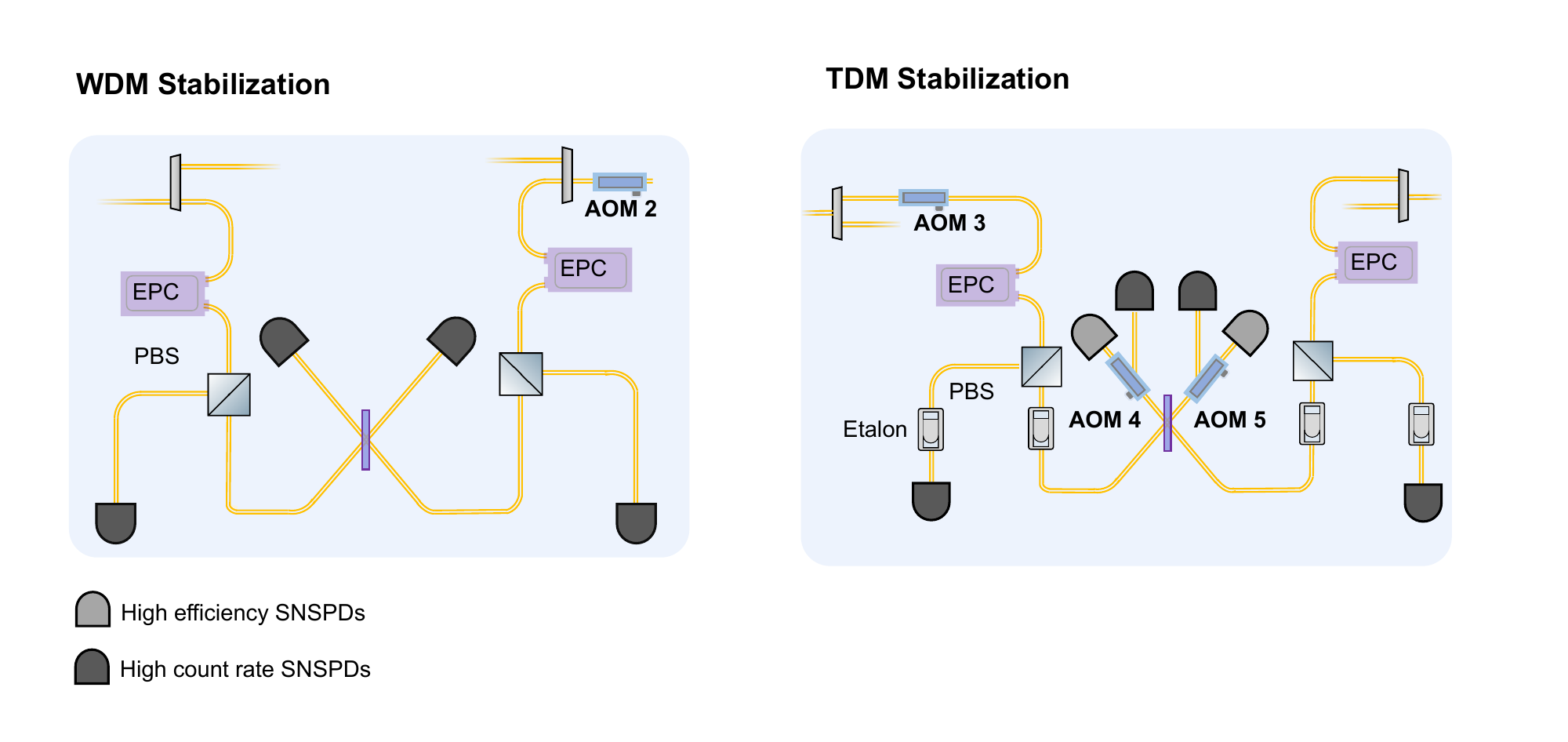}
\caption{\textbf{Schematic of the dual-band inter-node phase stabilization system architecture.} The hierarchical dual-stabilization scheme for inter-node phase coherence contains two independent channels. The left channel (WDM Stabilization) suppresses fast fiber link noise using a Polarization Beam Splitter (PBS), Electronic Polarization Controller (EPC), and superconducting nanowire single‐photon detectors (SNSPDs). The right channel (TDM Stabilization) compensates residual phase offset with an étalon, EPCs, and an AOM 4/AOM 5 cascade for switching between two SNSPDs: a high-efficiency detector for quantum-probe detection and a high-count-rate detector for reference-probe detection. Polarization components (PBS/EPC) enable purification. System coordination is performed by an FPGA controller using AOM 2/AOM 3 as core actuators for phase feedback.
}
\label{fig:feedback_sys}
\end{figure}

We implement wavelength-division multiplexed (WDM) stabilization for rapid fiber fluctuations and time-division multiplexed (TDM) stabilization for residual phase compensation. Phase information from both TDM and WDM channels is processed by an FPGA-based servo controller executing our prior-assisted estimation protocol at 100 kHz update rates. This system converts arrival-time-stamped photon counts into minimum-variance phase estimates and applies corrective feedback via AOMs strategically positioned in interferometer arms, as shown in Fig.~\ref{fig:feedback_sys}. The coordinated feedback achieves around 30 dB suppression of phase noise below 500 Hz. The high-efficiency SNSPD, featuring high quantum efficiency and low dark counts, is dedicated to quantum probe detection, while the high-count-rate SNSPD handles reference probe monitoring.

\subsection{FPGA-based real-time phase tracking system}

\begin{figure}[ht!]
\centering
\includegraphics[width=0.6\linewidth]{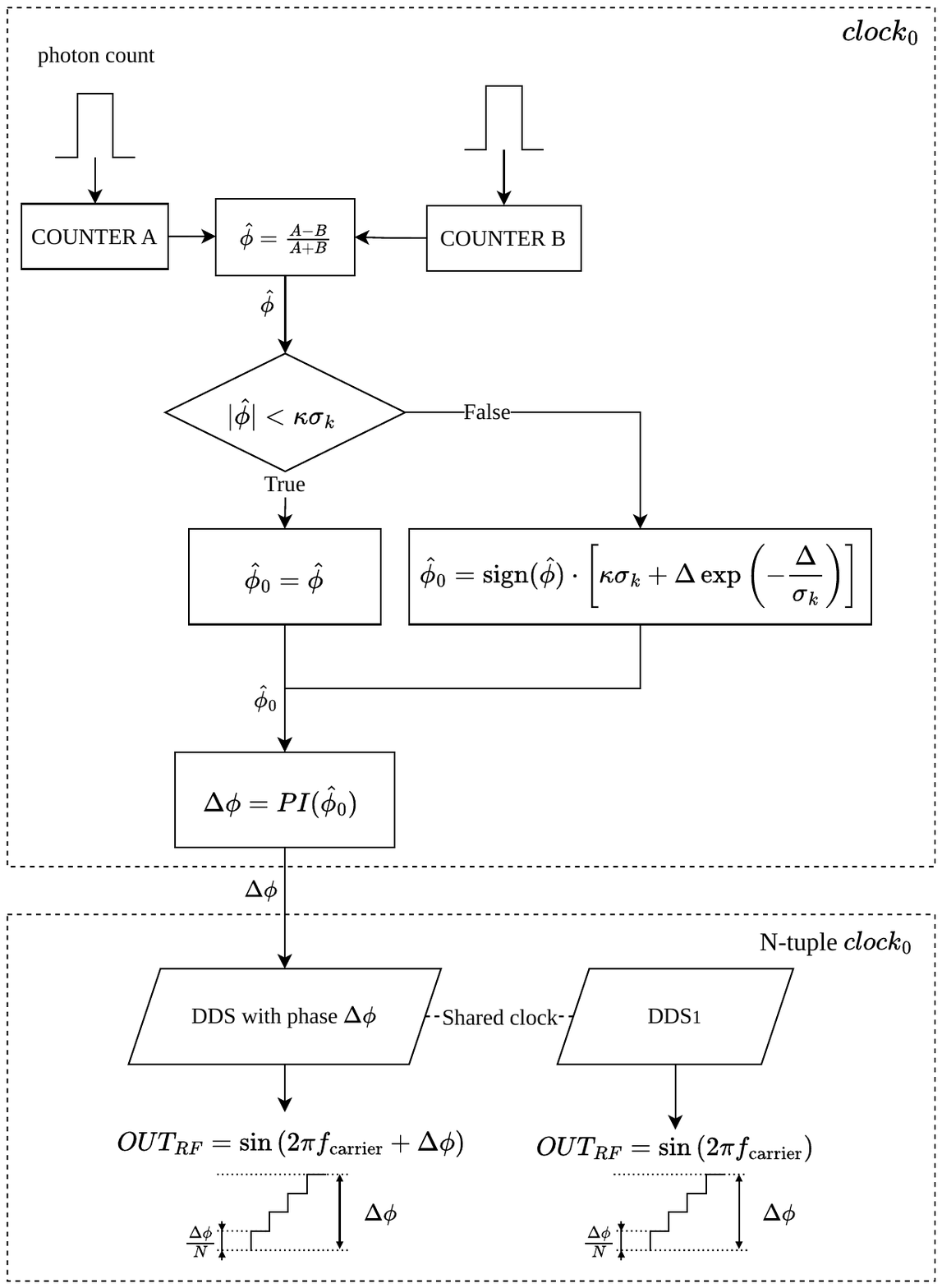}
\caption{\textbf{FPGA-based phase estimation and feedback control.} A schematic of the phase tracking system. Photon counts from dual interferometer outputs generate an error signal $\hat{\phi}$. A nonlinear innovation filter, employing recursive estimation with prior knowledge, conditions $\hat{\phi}$ to produce $\hat{\phi}_0$. A PI controller computes a phase correction. The entire process of detection, filtering, and PI control runs on $clock_0$. This correction value programs DDS modules, which operate at a higher frequency ($N\times clock_0$). This high-frequency operation applies corrections through multiple small increments per control cycle, reducing transient disturbances. The DDS modules generate the RF signals that actuate the AOMs, thereby closing the feedback loop for phase stabilization.}
\label{fig:Fig_FPGA_arg}
\end{figure}

A field-programmable gate array (FPGA) provides high-speed parallel processing for real-time phase tracking and feedback, enabling low-latency, high-precision phase stabilization. To achieve over 30 dB of low-frequency noise suppression, the system employs a sophisticated dual-AOM control architecture driven by an FPGA-based fast feedback algorithm. This architecture handles two perturbations in parallel: rapid, large-range phase drift from the fiber link and residual phase diffusion between interferometer arms. We adopt a dual-loop strategy in which AOM 2 rapidly corrects fiber-induced phase perturbations, while AOM 3 compensates the residual arm offset with higher precision. Each AOM is driven by an independent DDS. The following paragraphs detail the FPGA implementation of this control scheme.

The system implements a real-time phase-control algorithm on an FPGA to estimate and track optical phase shifts, the logical architecture is shown in Fig.~\ref{fig:Fig_FPGA_arg}. In this system, single-photon signals from the two output ports of the interferometer are first detected by high-efficiency SNSPDs and converted into electrical pulses. These pulses are discriminated and then fed into two independent counters within the FPGA for real-time counting.

The counter outputs are fed into a subtractor to generate an error signal, $\hat{\phi}$, representing the real-time phase difference between the two interferometer arms. This signal is processed by the nonlinear innovation filter in Eq.~\eqref{eq:filter}, which operates conditionally on the magnitude of $\hat{\phi}$ relative to the dynamic threshold $\kappa\sigma_k$ (where $\sigma_k$ is the prior-assisted estimate). Specifically, if $|\hat{\phi}| < \kappa \sigma_k$, the filter passes the signal unchanged ($\hat{\phi}_0 = \hat{\phi}$); otherwise, nonlinear filtering is applied with $\Delta = \hat{\phi} - \kappa\sigma_k$ quantifying the excess deviation beyond the statistically plausible bound. This design exponentially suppresses unreasonable outliers arising from shot-noise while preserving rapid response to small, legitimate error signals.

The shaped signal, $\hat{\phi}_0$, is then passed to a Proportional-Integral (PI) controller, which converts the error signal into a high-precision phase control word, $\Delta\phi$. The nonlinear mapping is precomputed and stored in a Look-Up Table (LUT). This optimization avoids real-time complex arithmetic, significantly conserves the FPGA's logical resources, and simultaneously guarantees real-time processing performance.

Finally, the phase control word, $\Delta\phi$, is written to a direct digital synthesizer (DDS), which generates a phase-modulated radio frequency (RF) signal, $  \text{OUT}_{\text{RF}} = \sin{(2\pi f_{\text{carrier}}+\Delta\phi)} $, to drive an AOM for phase correction. The DDS operates at high feedback speed, minimizing transient disturbances introduced by the correction process itself. Within the system, the feedback-controlled DDS and the fixed-frequency DDS1 (driving AOM 1) share a common clock reference. This architecture mitigates electronics-induced noise and relative clock drift, thereby ensuring long-term phase stability.

\subsection{Interferometric visibility characterization}

For the Mach-Zehnder interferometer configuration, the intensities at the two output ports are given by:
\begin{equation}
I_1(\phi) = \frac{N}{2}(1+V_0\cos\phi), \quad I_2(\phi) = \frac{N}{2}(1-V_0\cos\phi)
\end{equation}
where $\phi$ is the relative phase between the two 1550 nm probe signals, $N$ is the mean photon number, and $V_0$ is the intrinsic visibility. The phase tracking and feedback system locks the relative phase $\phi$ near $\pi/2$ by balancing the beam splitter (BS) port counts. 

For visibility measurements, we employ a time-division multiplexing sequence shown in Fig.~\ref{fig:Fig_parity_sequence}(c) to send the probe signal. We scan the relative phase between the two nodes' 729 nm beams to detect the probe interference fringe. Because the 1550 nm photon phase is inherited from the 729 nm and 854 nm, the 729 nm and 854 nm beams generate the probe signal through the PPLN waveguide. These 20 ns probe pulses copropagate along the same path as the emitted 393 nm single photons, thereby characterizing the phase stability of our ion-ion entanglement stabilization method. The probe signal is detected by SNSPDs, converted into electrical pulses, and recorded by a time-to-digital converter (TDC).

\begin{figure}[ht!]
\centering
\includegraphics[width=1\linewidth]{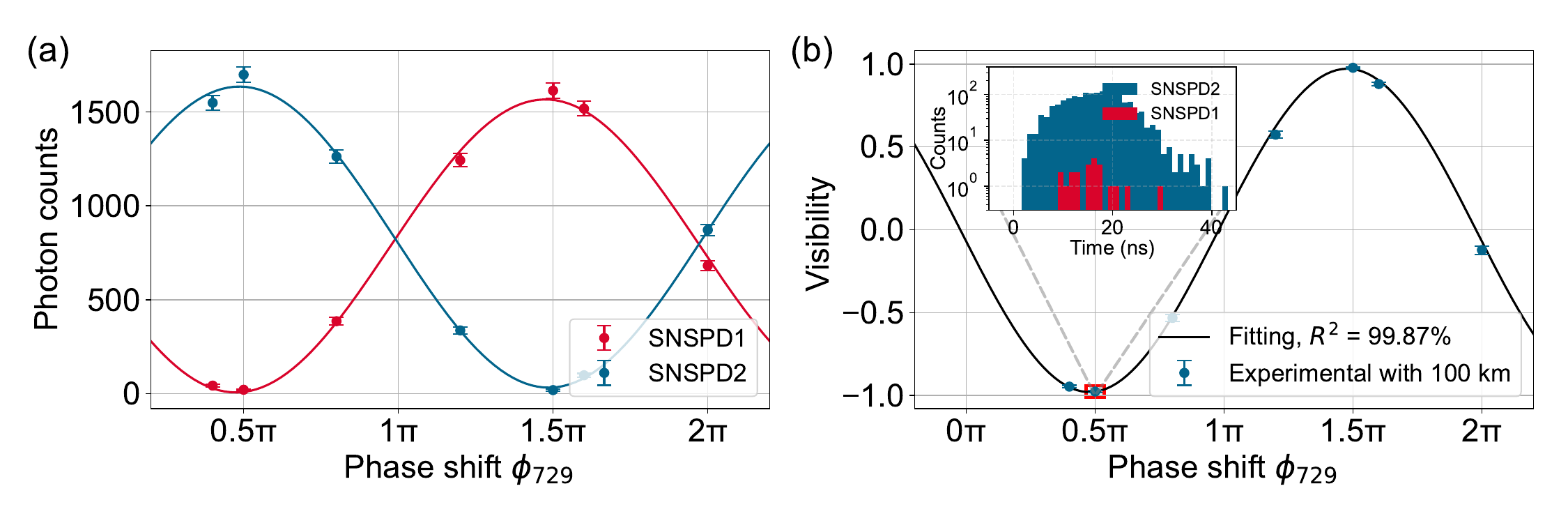}
\caption{\textbf{Phase-scanned interference fringes and visibility measurement.}
(a) Photon counts from the two SNSPD detectors (SNSPD1, SNSPD2) as a function of the scanned phase $\phi_{729}$ over a total measurement time of 100 seconds. The counts of SNSPD1 and SNSPD2 exhibit sinusoidal oscillations with similar amplitudes and opposite phases, indicating high interference contrast of the system.
(b) Visibility calculated from the data in (a) as a function of the phase $\phi_{729}$. The black solid line represents the fitting curve based on the formula Eq.~\eqref{eq:probe_vis} (goodness of fit $R^2 = 99.87\%$), and the blue dots represent experimental data obtained with a 100 km fiber. The fitted interferometer visibility is $V_0=97.49\%\pm0.66\%$.
(Inset in b) The inset shows the histogram  of detector counts at the phase-extreme point ($\pi/2$), showing a broadened pulse width ($\sim 40$ ns, due to etalon shaping) compared to the nominal 20 ns. Any minor phase perturbation will cause the operating point to deviate from the nominal phase value, leading to significant changes in the signal from the dark port and consequently a direct degradation of the measured visibility.}
\label{fig:visibility_char}
\end{figure}

During the measurement, the phase of the reference probe's 729 nm beam is held fixed. Thus, scanning the phase $\phi_{729}$ between the nodes' beams effectively scans the relative phase between the 20 ns probe pulses and the reference probe. The applied phase shift $\phi_{729}$ is inherited by the corresponding 1550 nm probe signal, establishing the final phase difference. Under phase-locked conditions, this scan produces sinusoidal modulation of the intensities at the two beam-splitter output ports, corresponding to alternating bright and dark fringes (Fig.~\ref{fig:visibility_char}(a)). Each specific phase setting $\phi_{(729,i)}$ yields a distinct ratio between the detector counts, from which the visibility $V_i$ is computed as:
\begin{equation}
    V_i = \frac{N_{\text{SNSPD1}}-N_{\text{SNSPD2}}}{N_{\text{SNSPD1}}+N_{\text{SNSPD2}}} = V_0\cos{(\pi/2+\phi_{(729,i)})}
\label{eq:probe_vis}
\end{equation}
$N_{\mathrm{SNSPD}1}$ and $N_{\mathrm{SNSPD}2}$ are the photon counts recorded by SNSPD1 and SNSPD2, respectively.
The intrinsic contrast $V_0$ is extracted by fitting the above model to the measured data points. The measurement is most sensitive near the fringe extrema, where even small phase noise changes the dark-port signal and reduces the measured visibility. At maximum interference contrast, phase instability reduces the visibility according to $V_{\text{max}}=V_0e^{-\sigma^2_{\phi}/2}$.

\subsection{ Interferometric visibility error source analysis}
\label{subsec:visibility_error}

The system-limited maximum visibility is $V_0=98.26\%\pm1.58\%$, constrained primarily by intrinsic optical imperfections, including electro-optic modulation distortion in the 854 nm excitation system and nonlinear noise in the frequency-conversion references. In contrast, the visibilities observed during actual entanglement-generation experiments are $97.98\% \pm 0.31\%$ for the 10-km link and $ 97.49\% \pm 0.66 \%$ for the 100-km link. The deficit between these operational values and $V_0$ arises from four real-time experimental factors rather than system limitations.

First, although the prior-assisted Bayesian estimator alleviates the precision--delay trade-off in the photon-starved regime, unavoidable feedback latency and measurement noise inevitably produce an irreducible phase residual that degrades visibility even under ideal conditions. Applying Eq.~(\ref{Eq:general_bay}) to the measured data with $\eta(\kappa)=1$ to estimate this residual-induced loss, we obtain $\sim 0.31\%$ for the 10\,km link and $\sim 0.44\%$ for the 100\,km link.

Second, temporal-overlap and pulse-shape mismatch impose a systematic penalty. The visibility-calibration probes are generated by electro-optic modulators that produce short (20 ns) optical pulses. The maximum achievable visibility depends critically on the temporal overlap and waveform matching between the probes arriving from Alice and Bob. Residual timing jitter, limited by the 100\,ps resolution of the time-to-digital converter (TDC) which introduces a $\sim$50\,ps uncertainty in determining the envelope alignment of the 40\,ns wavepackets (broadened by the etalon), together with pulse-shape asymmetry from modulator imperfections or fiber dispersion, further degrades the interference visibility. Additionally, intensity imbalance between the two wavepackets introduces an additional reduction in visibility. As shown in Fig.~3(c) of the main text, the visibility can reach $V_0$ under optimized conditions. During several-hour experimental runs, however, gradual timing-alignment drift causes the observed visibility to fluctuate within the 97--98\% range.

Third, polarization-dependent detection efficiency fluctuations in the SNSPDs introduce potential biases in both phase locking and visibility readout. Although the photons are coupled through short single-mode fibers ($\sim$ 3 m), residual polarization-dependent efficiency fluctuations($\sim$ 5\%)—typically on the order of a few percent over several hours—persist despite active polarization purification.  These slow drifts explain why the visibility varies between experiments and why achieving $V_0$ requires stable polarization and detector efficiency. We show below that the resulting locking and readout biases are negligible, respectively.
    
We first consider the effect on phase locking. Let $r=\eta_1/\eta_2$ denote the relative efficiency of the two detection channels, where $\eta_i$ is detector efficiency. The normalized error signal used for locking is $S=(I_{\text{SNSPD1}}-I_{\text{SNSPD2}})/(I_{\text{SNSPD1}}+I_{\text{SNSPD2}})$. Defining the normalized imbalance parameter $\xi=(1-r)/(1+r)$, the error signal near the lock point $\varphi=\pi/2+\delta_{\mathrm{lock}}$ linearizes to $S\approx \xi - V_{\text{0}}(1-\xi^2)\delta_{\mathrm{lock}}$. The feedback loop drives $S\to0$, yielding a lock-point offset of $\delta_{\mathrm{lock}}\approx \xi/V_{\text{0}}$. For typical parameters ($V_{\text{0}}=0.98$, $r=1.05$), this offset is merely $\sim1.4^\circ$. If this $5\%$ efficiency imbalance is treated conservatively as an rms fluctuation, it causes a visibility loss of only $\sim0.031\%$.

The second term acts during the visibility readout. In the presence of a static channel-efficiency imbalance, the measured normalized phase scan is not a pure sinusoid. The fitted visibility $V_{\rm fit}$ is the cosine amplitude extracted by the sinusoidal fit
\begin{equation}
V_{\rm fit}
=
\frac{1}{\pi}
\int_{0}^{2\pi}
S(\phi)\cos\phi\,d\phi =
-\frac{8r}{V_{\rm 0}(r-1)^2}
\left[
1-\frac{r+1}
{\sqrt{(r+1)^2-V_{\rm 0}^2(r-1)^2}}
\right].
\end{equation}
For $V_{\rm 0}=98.25\%$, the fitted-visibility bias is less than $0.017\%$.

Thus, a $5\%$ channel-efficiency fluctuation contributes about $0.031\%$ through lock-point jitter and less than $0.017\%$ through readout bias; even added conservatively, this is below $0.05\%$ and is not the dominant source of the operational visibility reduction.

Finally, the finite servo bandwidth of the time-division multiplexing stabilization loop limits phase coherence. The feedback operates only during brief phase-probe windows (65\,$\mu$s per 500\,$\mu$s cycle for 10\,km; 95\,$\mu$s per 800\,$\mu$s cycle for 100\,km), relying on a predictive model between windows. This intermittent operation reduces the effective loop gain and update rate. The dominant phase noise from the 20 m fibers is concentrated below 500 Hz, within the servo bandwidth even at the reduced update rate. This bandwidth limitation is tied to the heralding protocol: increasing the probe duty cycle would require shorter ion-cooling and manipulation sequences, which would degrade the entanglement fidelity. The 6.5\% duty cycle represents the optimal trade-off between phase stabilization and quantum state operation.

The reported operational visibilities represent routinely achievable performance under standard operating conditions without fine-tuning. In dedicated calibration runs with optimized parameters, the observed visibility can reach the system-limited upper bound $V_0$, supporting the conclusion that the gap arises from the real-time factors described above rather than from irreversible system defects.

\begin{table}[t]
\caption{Visibility error budget for the 10 km and 100 km links.}
\centering
\label{tab:visibility_error_budget_simple}
\begin{tabular}{lcc}
\hline
Error source & 10-km link & 100-km link \\
\hline
Bayesian residual phase variance &
$\approx$ 0.31\% &
$\approx$ 0.44\% \\
Timing / waveform jitter &
\(< 0.12\)\% &
\(< 0.12\)\% \\
Detector efficiency fluctuations  &
\(<0.05\)\% &
\(<0.05\)\% \\
The finite servo bandwidth   &
\(\sim0.01\)\% &
\(<0.20\)\% \\
\hline
Total estimated contribution &
\(< 0.49\)\% &
\(< 0.81\)\% \\
Measured deficit from \(V_0\) &
\(0.28\)\% &
\(0.77\)\% \\
\hline
\end{tabular}
\end{table}

The error analysis above pertains to a laboratory demonstration where the 729\,nm, 854\,nm, and 527\,nm lasers at both Alice and Bob are locked to the ultra-stable cavities, ensuring negligible relative frequency offsets between corresponding wavelengths. In field deployment, however, Alice and Bob operate with independent lasers connected via optical fiber. While the WDM architecture suppresses fiber-induced phase noise and the 20\,m spool faithfully simulates the phase fluctuations of independent cavity-locked lasers, a slow relative frequency drift of several to tens of Hz persists due to differential temperature drifts between the remote cavities. Characterizing visibility degradation under such frequency offsets is therefore essential for validating system performance in realistic operational scenarios.

To evaluate interferometric-visibility degradation under independent laser operation, we applied a controlled relative frequency offset between the 1550\,nm light at Alice and Bob, thereby emulating independent local oscillators. We then measured the interference visibility as a function of this offset. The optical frequency of the 1550\,nm light is determined by the 729\,nm, 854\,nm, and 527\,nm beam frequencies through cascaded nonlinear frequency conversion
\begin{equation}
    \nu_{1550}=\nu_{729}+\nu_{854}-\nu_{527},
\end{equation}
where $\nu_i$ denotes the optical frequency of the $i$\,nm beam. Thus, the relative frequency offset between the 1550\,nm light at Alice and Bob can be controlled by adjusting the AOM drive frequency in either the 729\,nm or 854\,nm path at Alice. We measured the interference visibility under three configurations: offset applied solely to the 729\,nm beam, solely to the 854\,nm beam, and simultaneously to both. Single-beam offset test points were set at +30, +50, +100, +300, and +500\,Hz, while the joint-offset configuration applied +50\,Hz to each beam (yielding a net offset of +100\,Hz).

\begin{figure}[ht!]
\centering
\includegraphics[width=0.5\linewidth]{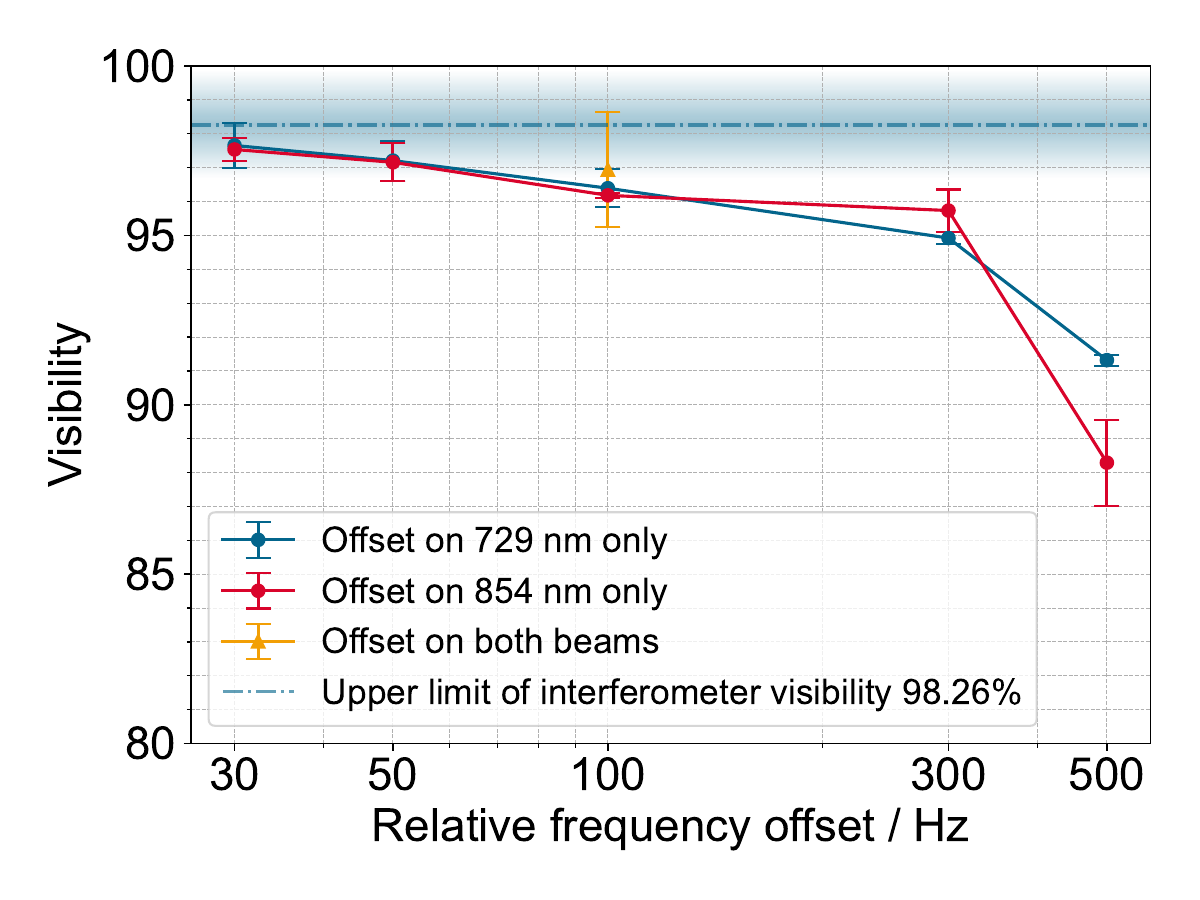}
\caption{\textbf{Effect of frequency offset on interferometric visibility.} Visibility of the 1550\,nm light generated via cascaded nonlinear frequency conversion as a function of the relative frequency offset between Alice and Bob. Blue, red, and yellow markers correspond to offsets applied solely to the 729\,nm beam (Offset on 729\,nm only), solely to the 854\,nm beam (Offset on 854\,nm only), and jointly to both beams (Offset on both beams), respectively. For the joint configuration, $+50$\,Hz is applied to each beam, yielding a net offset of $+100$\,Hz. The visibility degrades monotonically with increasing offset across all configurations, remaining above 97\% for offsets $\leq$\,50\,Hz. The blue dash-dotted line indicates the upper limit of the interferometer visibility.}
\label{fig:frequency_offset}
\end{figure}

The measured visibility is identical for single-beam and joint-offset cases with the same net offset, indicating that the conversion process is insensitive to the source of the perturbation. The visibility depends solely on the relative frequency difference between the 1550\,nm light. As shown in Fig.~\ref{fig:frequency_offset}, the visibility remains above 97\% for offsets $\leq$\,50\,Hz and degrades to a minimum of $96.18\% \pm 0.08\%$ across all three configurations at 100\,Hz. This offset exceeds the typical relative frequency fluctuation of our ultra-stable cavity-locked laser systems, which remains far below 100\,Hz over the experimental timescale.

\subsection{Parity contrast measurements and error source analysis}
\label{subsec:parity_sensitivity}

Parity detection allows the phase sensitivity to be computed directly from the expectation value $\langle \hat{\Pi}\rangle$ and its derivative, without post-processing or reconstruction of probability distributions, unlike photon-counting strategies~\cite{PhysRevA.87.043833}. This makes parity measurement particularly attractive for real-time, high-precision phase estimation in quantum metrology. The parity is the key metric for entangled-state fidelity. For the maximum entangled state $|\Psi_{+}\rangle = \frac{1}{\sqrt{2}}(|SD\rangle+i|DS\rangle)$, the fidelity is
\begin{equation}
\mathcal{F}=\langle \Psi_+| \rho^{\text{exp}}|\Psi_+\rangle = \frac{1}{2}(\rho_{SD,SD}^{\text{exp}}+\rho_{DS,DS}^{\text{exp}})+\text{Im}\,\rho_{SD,DS}^{\text{exp}}.
\end{equation}

\begin{figure}[ht!]
\centering
\includegraphics[width=\linewidth]{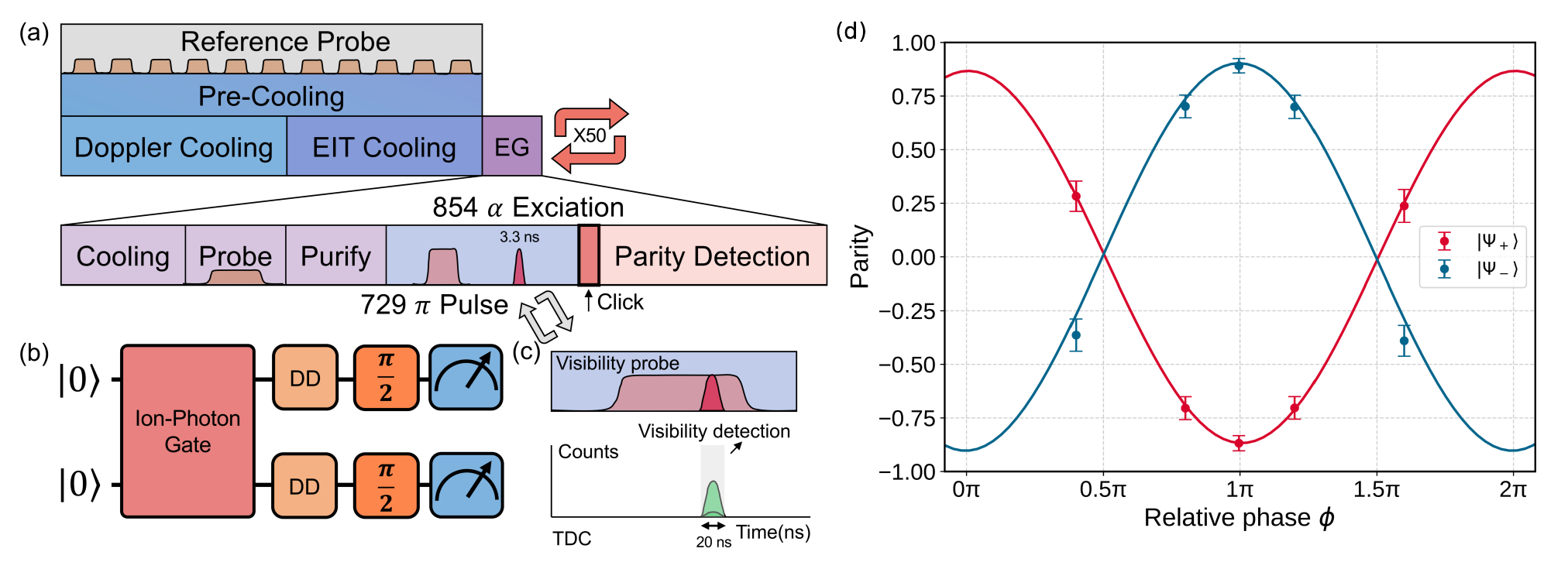}
\caption{\textbf{Heralded entanglement generation and Bell-state verification via parity oscillation.}
(a) Overall experimental sequence. Following the pre-cooling stage with reference probe pulses for phase stabilization, the entanglement generation (EG) inner loop is repeated 50 times. Each EG cycle includes: ion cooling to reduce atomic recoil errors, a reference probe sequence (applied every 65~$\mu$s for the 10~km configuration and every 95~$\mu$s for the 100~km configuration), purification to prepare the $|S_{1/2}, m_j = +1/2\rangle$ state, and single-photon generation via a 729~nm $\pi$ pulse followed by a nanosecond-scale 854~nm $\alpha$ excitation pulse, which emits a 393~nm photon frequency-converted to 1550~nm. Upon detection of a photon click at the mid-station by the SNSPD, heralding successful ion-ion entanglement, parity detection is performed.
(b) Quantum circuit for parity detection. Following the heralded entanglement (represented by the ion-photon gate), dynamical decoupling (DD) pulses are applied to extend the coherence time, followed by $\pi/2$ analysis pulses with variable phase to measure the parity operator $\langle \sigma^{(1)}_\phi \sigma^{(2)}_\phi \rangle$ via 397 nm fluorescence detection.
(c) Visibility measurement sequence. The 729~nm and 854~nm beams are overlapped temporally to generate 1550~nm photons. The interference signal is detected by SNSPDs and recorded by a time-to-digital converter (TDC). (d) Measured parity $\langle \sigma^{(1)}_\phi \sigma^{(2)}_\phi \rangle$ as a function of relative phase $\phi$ for the two maximally entangled Bell states $|\Psi_+\rangle = \frac{1}{\sqrt{2}}(|01\rangle + i|10\rangle)$ (red circles) and $|\Psi_-\rangle = \frac{1}{\sqrt{2}}(|01\rangle - i|10\rangle)$ (blue circles), with excitation probability $\alpha = 5\%$. The solid curves are sinusoidal fits $P_{\text{fit}}(\phi) = A\sin(2\phi+\phi_0)$, yielding fringe amplitudes of $A = 0.867\% \pm 0.008\%$ for $|\Psi_+\rangle$ and $A = 0.903\% \pm 0.005\%$ for $|\Psi_-\rangle$. Error bars represent one standard deviation from photon counting statistics.} 
\label{fig:Fig_parity_sequence}
\end{figure}

Here, $\rho^{\text{exp}}$ is the density matrix of the experimentally produced entangled state. To characterize the entanglement, we measure off-diagonal elements using measurements~\cite{sackett2000experimental} $P(\phi)= \langle \sigma^{(1)}_{\phi}\sigma^{(2)}_{\phi}\rangle$ for varying phase $\phi$, where $\sigma_\phi = \sigma_x\cos\phi+\sigma_y\sin\phi$. The measured parity oscillation $P(\phi)$ is fitted to $P_{\text{fit}}(\phi) = A\sin(2\phi+\phi_0)$, as shown in Extended Fig.~\ref{fig:Fig_parity_sequence}(d).

The quantum circuit for parity detection is shown in Fig.~\ref{fig:Fig_parity_sequence}(b). After generating the ion-photon entangled state and detecting a photon click at the mid-station to herald ion-ion entanglement, we perform parity measurements by applying a local $\pi/2$ pulse after the dynamical decoupling sequence and scanning its phase $\phi$. This method is analogous to Ramsey interference measurements between the $|SD\rangle$ and $|DS\rangle$ states. The complete experimental sequence is illustrated in Fig.~\ref{fig:Fig_parity_sequence}(a).

We next consider the dominant error sources in the parity measurement.

\begin{table}[ht!]
\centering
\caption{\textbf{Estimated parity-error budget for remote ion-ion entanglement generation.}}
\begin{tabular}{|m{4cm}|m{3cm}|m{8cm}|}
\hline
\multicolumn{1}{|c|}{\textbf{Error Source}} & \multicolumn{1}{c|}{\textbf{Contribution}} & \multicolumn{1}{c|}{\textbf{Description}} \\
\hline
\hline
\parbox[c][1.5cm][c]{4cm}{\centering Phase Stability} & \parbox[c][1.5cm][c]{3cm}{\centering $\epsilon_{\text{phase stab}}<2\%$} & 
\parbox[c][1.5cm][c]{8cm}{Residual phase instability at $\theta=\pi/2$ reduces parity contrast as 
$P =\langle e^{i\delta\theta} \rangle =e^{-\sigma_{\theta}^2/2}$} \\
\hline
\parbox[c][1.5cm][c]{4cm}{\centering Atomic Motion \& Micromotion} & \parbox[c][1.5cm][c]{3cm}{\centering $\epsilon_{\text{motion}}<3\%$} & 
\parbox[c][1.5cm][c]{8cm}{Ion motion modulates photon phase, reducing contrast via 
$P_{\text{motion}} = V_0J_0(2kA_m)e^{-2(k\sigma)^2} $, where $k$ is photon wave vector and $\sigma$ is radial motional amplitude} \\
\hline
\parbox[c][1.5cm][c]{4cm}{\centering Single-Photon Protocol} & \parbox[c][1.5cm][c]{3cm}{\centering $\alpha\le 5\%$} & 
\parbox[c][1.5cm][c]{8cm}{Weak excitation probability in ion-photon state 
$|\Psi\rangle = \sqrt{1-\alpha}|D,0\rangle+\sqrt{\alpha}e^{i\phi}|S,1\rangle$ leads to intrinsic parity reduction} \\
\hline
\parbox[c][1.5cm][c]{4cm}{\centering Qubit Manipulation} & \parbox[c][1.5cm][c]{3cm}{\centering $\epsilon_{\text{manipulation}}<1\%$} & 
\parbox[c][1.5cm][c]{8cm}{Imperfect $\pi$ rotation ($\epsilon_{\text{rot}}<0.4\%$) and SPAM errors ($\epsilon_{\text{SPAM}}<0.2\%$)} \\
\hline
\parbox[c][1.5cm][c]{4cm}{\centering Memory Decoherence} & \parbox[c][1.5cm][c]{3cm}{\centering $\epsilon_{\text{deco}}<2\%$} & 
\parbox[c][1.5cm][c]{8cm}{Coherence loss during photon propagation (25~$\mu s$ for 5~km, 250~$\mu s$ for 50~km) plus classical signal return (25~$\mu s$ or 250~$\mu s$), totaling 50~$\mu s$ (500~$\mu s$)} \\
\hline
\parbox[c][1.5cm][c]{4cm}{\centering Link Noise} & \parbox[c][1.5cm][c]{3cm}{\centering $\epsilon_{\text{link}}<1\%$} & 
\parbox[c][1.5cm][c]{8cm}{QFC noise photons and SNSPD dark counts create false heralding events, preparing erroneous state $|\Psi_{\text{err}}\rangle=|SS\rangle$ with parity error $1/(\text{SNR}+1)$} \\
\hline
\end{tabular}
\label{tab:error_sources}

\end{table}

After accounting for these error sources, the expected parity fringe contrast exceeds $86\%$. The measured parity values are $88.4\%\pm5.1\%$ for the 10 km link and $87.3\%\pm6.5\%$ for the 100 km link, in agreement with the contrast predicted from the dominant error contributions.

\subsection{Experimental sequence for parity and visibility measurements}

To reduce atomic recoil errors, we employ pre-Doppler cooling followed by pre-EIT cooling to prepare both radial and axial motional modes near ground states. After cooling, the two radial modes are cooled to $\bar{n} \approx 0.5$ and the axial mode to $\bar{n} \approx 1$. However, the reference probe emits 393 nm light for 65 $\mu$s every 500 $\mu$s 
(95~$\mu$s per 800~$\mu$s for 100~km) to maintain phase stability, which can perturb the qubit state. To protect the qubit during the pre-EIT cooling stage, we shelve the population to the $D_{3/2}$ state using a 3~$\mu$s, 397~nm pulse, then restore it to $|S\rangle$ using a 3~$\mu$s, 866~nm beam after cooling. Although this shelving process introduces additional motional heating, the final mean phonon numbers remain low enough to maintain high parity contrast.

Each entanglement generation attempt consists of three steps: (1) State preparation to $|S_{1/2}, m_j = +1/2\rangle$ via optical pumping with 729~nm and 854~nm beams; (2) Single-photon generation using a resonant 729~nm $\pi$ pulse to coherently drive the $|S_{1/2},m_j=+1/2\rangle \to |D_{5/2},m_j=+5/2\rangle$ transition, followed by a short 854~nm excitation pulse to generate the 393~nm photon; (3) Waiting for the heralding signal from mid-station detection, followed by parity analysis using the rotation-scan protocol described above. To avoid ion non-crystallization~\cite{PhysRevA.105.033101}, we repeat the entanglement-generation attempt up to 50 times.

For visibility measurements, the time-division multiplexing sequence is nearly identical to the parity-measurement sequence, except that visibility measurements do not require precooling, state preparation, or parity-measurement operations. Both output ports of the Mach--Zehnder interferometer are detected by SNSPDs and recorded via the TDC. During each 393~nm probe emission, we set the 729~nm beam phase to $\phi$, then reset it to $\phi_0$ for the subsequent phase stabilization cycle. By scanning $\phi$ and measuring the interference fringe at both output ports, we extract the visibility according to Eq.~\eqref{eq:probe_vis}.

\subsection{Motion-induced parity loss from recoil and micromotion}

Atomic motion reduces the parity contrast of the heralded ion-ion state through recoil-induced coupling between the emitted photon and the ion's motional states. This effect is distinct from optical phase-stabilization error. The phase lock determines the mean Bell-state phase, whereas residual motion reduces the coherence with which this phase is defined after tracing over the unobserved motional state. Following the kick-operator description of photon-mediated entanglement~\cite{apolin2026recoil,saha2025high}, photon emission from ion $q$ applies the motional recoil operator

\begin{equation}
    \hat K_q=\exp(i\Delta\mathbf{k}_q\cdot\hat{\mathbf r}_q),
\end{equation}
where $\hat{\mathbf r}_q$ is the ion position operator. After factoring out the common 729 nm preparation pulse, the recoil is dominated by the 854 nm excitation and the collected 393 nm emission. For motional mode $j$, the effective Lamb-Dicke parameter is
\begin{equation}
    \eta_{qj}
    = (\Delta\mathbf{k}_q\cdot\mathbf e_{qj})
    \sqrt{\frac{\hbar}{2m\omega_{qj}}}.
\end{equation}
Tracing over a thermal motional state gives the single-node coherence reduction
\begin{equation}
    C_q^{\rm wp}
    =
    \exp\!\left[
    -\sum_j
    \left(\bar n_{qj}+\frac12\right)\eta_{qj}^2
    \right].
\end{equation}
For the heralded two-ion state, the Bell-state coherence, and hence the parity contrast, is multiplied by $C_{\rm wp}=C_A^{\rm wp}C_B^{\rm wp}$, so that the thermal-wavepacket recoil contribution to the parity-contrast loss is $\epsilon_{\rm wp}^{(P)}=1-C_{\rm wp}$.

The measured secular frequencies and phonon occupations used in this estimate are listed in Table~\ref{tab:motion_frequencies_photon}.

\begin{table}[h]
\centering
\caption{Measured secular frequencies and motional phonon used for the motion-induced parity-loss estimate. Frequencies are given in MHz. The phonon occupations are measured after the relevant wait time in the entanglement-generation sequence.
}
\begin{tabular}{lcccc}
\hline\hline
Node & Mode & \(\omega_{qj}/2\pi\) (MHz) & \(\bar n\) (10 km) & \(\bar n\) (100 km) \\
\hline
Alice & Radial 1 & \(1.6489\) & \(0.92\pm0.08\) & \(0.60\pm0.03\) \\
Alice & Radial 2 & \(2.3590\) & \(0.81\pm0.06\) & \(0.66\pm0.07\) \\
Alice & Axial    & \(0.8759\) & \(1.3\pm0.4\)   & \(1.0\pm0.3\) \\
\hline
Bob   & Radial 1 & \(1.7745\) & \(0.86\pm0.05\) & \(0.81\pm0.10\) \\
Bob   & Radial 2 & \(2.1415\) & \(0.75\pm0.08\) & \(0.95\pm0.14\) \\
Bob   & Axial    & \(1.0678\) & \(2.6\pm0.7\)   & \(2.8\pm0.4\) \\
\hline\hline
\end{tabular}
\label{tab:motion_frequencies_photon}
\end{table}

We also account for the finite collection cone of the $\rm NA=0.635$ objective. This yields $\epsilon_{\rm wp}^{(P)}\lesssim2.9\%$ for the 10 km and 100 km sequence. 
Using the 729 nm sideband spectrum, we take the two radial modes to have comparable projection onto the 729/854 beam direction, $u_{r1}\simeq u_{r2}\simeq1/2$, while the axial mode is nearly orthogonal, $u_z\simeq0$, where $u_j=(\mathbf e_j\cdot\hat{\mathbf z})^2$. The NA-averaged recoil projection is then
\begin{equation}
\begin{split}
\left\langle
(\Delta\mathbf{k}\cdot\mathbf e_j)^2
\right\rangle_{\rm NA}
=&\;
u_j
\left[
k_{854}^2
-2k_{854}k_{393}\langle\cos\theta\rangle
+k_{393}^2\langle\cos^2\theta\rangle
\right] \\
&+(1-u_j)k_{393}^2
\frac{1-\langle\cos^2\theta\rangle}{2}.
\end{split}
\end{equation}
This yields $\epsilon_{\rm wp}^{(P)}\lesssim2.9\%$ for the 10 km and 100 km sequence.

A second, smaller contribution comes from averaging over the finite temporal extent of the emitted photon wavepacket and the accepted detection window. As discussed in Ref.~\cite{saha2025high}, different emission or detection times correspond to slightly different recoil displacements in motional phase space. In the limit \(\omega_{qj}\tau_R\ll1\), the two-node contrast factor is
\begin{equation}
    C_{\rm win}
    =
    \prod_{q=A,B}\prod_j
    \exp\!\left[
    -\left\langle\zeta_{qj}^2\right\rangle_{\rm NA}
    (2\bar n_{qj}+1)W\omega_{qj}^2\tau_R^2
    \right],
\end{equation}
where $W\le1$ is determined by the accepted detection window. The NA-averaged
emission recoil is
\begin{equation}
    \left\langle\zeta_{qj}^2\right\rangle_{\rm NA}
    =
    \frac{\hbar k_{393}^2}{2m\omega_{qj}}
    \left[
    u_j\langle\cos^2\theta\rangle
    +(1-u_j)\frac{1-\langle\cos^2\theta\rangle}{2}
    \right].
\end{equation}
Taking the conservative limit \(W=1\) and summing over the three motional modes
of both ions gives
\begin{equation}
    \epsilon_{\rm win}^{(P)}<0.07\%.
\end{equation}

Finally, residual coherent micromotion produces an additional phase modulation
of the emitted photon. Following the interferometric visibility treatment of a
single sub-Doppler-cooled ion~\cite{slodivcka2012interferometric}, the
RF-phase-averaged reduction is described by a Bessel factor. In our notation,
\begin{equation}
    C_{\rm mm}=\prod_{q=A,B}\prod_j J_0(2\beta_{qj}).
\end{equation}
Our micromotion compensation gives \(\beta_{qj}<0.005\) along all three
directions. Using \(J_0(2\beta)\simeq1-\beta^2\), we obtain
\begin{equation}
    \epsilon_{\rm mm}^{(P)}
    =
    1-C_{\rm mm}
    <
    0.015\%
\end{equation}
corresponding to a parity-contrast loss of $<0.015\%$. The total motion-induced parity contrast loss is therefore
\begin{equation}
    C_{\rm motion}=C_{\rm wp}C_{\rm win}C_{\rm mm},
    \qquad
    \epsilon_{\rm motion}^{(P)}=1-C_{\rm motion}.
\end{equation}
The resulting budget is summarized in Table~\ref{tab:motion_error_budget}.

\begin{table}[h]
\centering
\caption{Motion- and recoil-induced parity error budget.
}
\begin{tabular}{lc}
\hline\hline
Contribution & Estimate \\
\hline
Thermal-wavepacket recoil overlap error & $\lesssim2.9\% $ \\
Photon-arrival-time averaging error & \(<0.07\%\) \\
Residual micromotion error & \(<0.015\%\) \\
\hline
Total motion-induced parity loss & \(\lesssim 3.2\%\) \\
\hline\hline
\end{tabular}
\label{tab:motion_error_budget}

\end{table}

\subsection{Single-mode fiber coupling for photon interference}

The emitted photons are coupled into low-loss single-mode fiber (SMF) to define a common spatial mode for interference at Charlie. Although the selected atomic transition emits $\sigma_+$-polarized photons, high-NA collection samples a finite solid angle over which the local polarization basis of the dipole radiation varies~\cite{nadlinger2022a}. The collected mode therefore does not define a fixed laboratory-frame linear-polarization axis that can be directly matched to a polarization-maintaining fiber (PMF) eigenaxis.

Moreover, PMF can introduce additional temperature- and stress-sensitive differential phase shifts between polarization components~\cite{stephenson2020high}. This sensitivity arises from fiber birefringence
\begin{equation}
    \Delta \phi = \frac{2\pi}{\lambda}\Delta n L,
\end{equation}
where $\lambda$ is the optical wavelength, $\Delta n$ is the birefringence, and $L$ is the fiber length. Because $\Delta n$ is typically $\sim 10^{-4}$ for PMF, compared with $\sim 10^{-7}$ for standard SMF, even small temperature- or stress-induced fiber variations can produce appreciable differential phase shifts, thereby degrading photon indistinguishability. We therefore use standard SMF for low-loss spatial-mode filtering and long-distance telecom transmission, while polarization is actively aligned and purified after fiber coupling using an EPC and PBS before interference.

\bibliography{SM}